\documentclass[12pt]{article}

\usepackage{lineno}
\usepackage{tikz}
\usetikzlibrary{arrows.meta,positioning,shapes.geometric}
\usepackage{rotating}
\usepackage{longtable}
\usepackage{booktabs}
\usepackage{array}      
\usepackage{caption}
\usepackage{multirow}
\usepackage{tabularx}
\usepackage{placeins}
\usepackage{soul}
\newcolumntype{P}[1]{>{\raggedright\arraybackslash}p{#1}}

\usepackage{newtxtext,newtxmath}

\usepackage{graphicx}

\usepackage[letterpaper,margin=1in]{geometry}

\renewenvironment{abstract}
	{\quotation}
	{\endquotation}

\date{}

\makeatletter
\renewcommand{\fnum@figure}{\textbf{Figure \thefigure}}
\renewcommand{\fnum@table}{\textbf{Table \thetable}}
\makeatother

\usepackage{scicite}

\usepackage{url}

\usepackage{bbold}

\def\scititle{
    Remote work expands pathways to upward career mobility
}

\title{\bfseries \boldmath \scititle}

\author{
	Yunhan~Zheng$^{1,2\ast}$,
	Jinhua~Zhao$^{3}$\and
	\small$^{1}$College of Urban and Environmental Sciences, Peking University, Beijing 100871, PR China.\and
	\small$^{2}$Laboratory for Earth Surface Processes of the Ministry of Education, Peking University, Beijing 100871, PR China.\and
    \small$^{3}$Department of Urban Studies and Planning, Massachusetts Institute of Technology, Cambridge 02139, MA, USA.\and
	\small$^\ast$Corresponding author. Email: yunhan@pku.edu.cn
}

\begin{document} 

\maketitle

\begin{abstract} \bfseries \boldmath

Geographic constraints have long structured access to high-growth career opportunities, concentrating upward mobility within a limited set of cities and organizations. The expansion of remote work potentially alters this opportunity structure by decoupling job matching from physical proximity, yet its implications for career mobility remain unclear. Using 48 million U.S. job transitions between 2020 and 2024 linked to employer-level measures of remote eligibility, we estimate how entering remote-eligible jobs shapes career outcomes at job transitions. Workers entering remote-eligible jobs experience significantly higher wage growth and higher rates of upward seniority mobility than comparable workers entering fully on-site roles. These transitions are also associated with greater cross-metropolitan job mobility and moves toward smaller, less prestigious employers. Importantly, effects are largest among lower-income workers and those originating from regions with limited high-skill opportunity density. Together, the findings indicate that remote work relaxes geographic constraints in job matching, reshaping the distribution of upward mobility across places and workers.
\end{abstract}

\noindent
High-wage employment in the United States has become increasingly concentrated in a small set of highly productive metropolitan areas and dominant firms, often described in the literature as “superstar cities” and “superstar organizations”~\cite{kemeny2020superstar,autor2020fall,kemeny2024changing}. This spatial and organizational clustering has been argued to restrict access to upward career mobility. To enter high-growth career paths, workers must often relocate to expensive urban centers and compete for positions within a narrow set of selective employers~\cite{le2019superstar,abel2019some}. As a result, many qualified workers remain disconnected from the labor market’s most dynamic opportunities simply because they are geographically or institutionally out of reach~\cite{kemeny2020superstar,acolin2017opportunity,ishimaru2025geographic}.

The rapid expansion of remote work after the COVID-19 pandemic has begun to loosen these long-standing constraints. Remote and hybrid arrangements, now common among college-educated workers in the United States~\cite{bloom2024hybrid,aksoy2025working}, have reduced the extent to which career advancement depends on geographic proximity to economic hubs. In doing so, remote work potentially reshapes the geography of career opportunity by allowing workers to access employers beyond their local labor markets without requiring relocation or daily commuting to major economic hubs. This shift raises a central question for researchers and policymakers: does remote work genuinely broaden access to high-growth employment, or does it introduce new forms of disadvantage for those who adopt it?

Existing research highlights potential trade-offs associated with remote flexibility.
 The theory of compensating differentials emphasizes that workers may accept lower wages in exchange for valued amenities such as flexibility~\cite{mas2017valuing,barrero2021working,he2021workers}. Related work in urban and organizational economics underscores the role of spatial proximity and in-person interaction in shaping career pathways through firms and local labor markets \cite{glaeser2001cities,storper2004buzz,yang2022effects}. Yet despite these theoretical perspectives, empirical evidence on how remote work reshapes career trajectories remains limited. Existing studies—relying largely on cross-sectional surveys or single-firm case studies—have primarily focused on short-term outcomes, such as productivity or subjective well-being~\cite{bloom2024hybrid,gibbs2023work,aksoy2023time}. We lack systematic, economy-wide evidence on how remote work eligibility shapes long-term career outcomes, including wage outcomes and upward seniority mobility at job transitions, as well as patterns of organizational and geographic mobility.



To examine this question, we assemble a large-scale dataset that links workers’ career trajectories to the remote-eligibility status of the jobs they enter. We combine longitudinal employment histories covering tens of millions of job-entry events with hundreds of millions of job-posting records that indicate whether specific roles were designated as remote-eligible at the time of hiring. Importantly, remote-work eligibility is measured from employer job postings prior to the worker’s job transition, allowing us to characterize remote eligibility as a feature of job structure determined by employers rather than by individual negotiation or worker preferences. This linkage allows us to measure remote-work eligibility at the level of each job transition as a supply-side feature of the job determined by employers. We then compare workers entering remote-eligible and fully in-office positions with similar pre-entry characteristics and estimate how remote eligibility shapes subsequent career outcomes, including wage growth, upward seniority mobility, and patterns of organizational and geographic mobility.

Our findings suggest that remote work expands pathways to upward mobility that are not confined to elite organizations. Workers entering remote-eligible jobs experience higher wage growth and greater upward seniority mobility, accompanied by a distinctive pattern of organizational mobility and expanded access to jobs beyond local labor markets. Rather than primarily moving into prestigious organizations, workers frequently transition toward smaller and less prestigious employers, including those outside their local labor markets. These effects persist over time and vary systematically by workers’ prior earnings, with larger wage gains among lower-paid workers.

Our study makes three contributions to the literature on remote work and labor-market opportunity. First, using large-scale observational data on career histories and job postings, we provide systematic, economy-wide evidence that eligibility for remote work is associated with upward career mobility, rather than systematic career penalties. This evidence speaks to long-standing concerns about whether workplace flexibility comes at the expense of career advancement or wages\cite{mas2017valuing,goldin2014grand,barrero2021working}.

Second, we identify a distinctive pattern of upward mobility characterizing remote-eligible job transitions.
We find that workers entering remote-eligible jobs achieve higher wage growth and higher rates of upward seniority moves while sorting into organizations that are smaller and less prestigious.
This pattern departs from the conventional view that career advancement is primarily tied to employment at large, prestigious organizations, as emphasized in models of labor-market sorting \cite{card2013workplace,lazear1981rank}, revealing a distinct configuration of career advancement in remote-eligible labor-market transitions.

Third, while prior work emphasizes that access to remote work is disproportionately concentrated among high-skilled, high-income workers—raising concerns that remote work may exacerbate labor market inequality \cite{dingel2020many,barrero2021working}—we show that, after accounting for differences in workers’ prior earnings, remote-eligible job entry is associated with higher wage growth and upward seniority mobility across the earnings distribution.
These gains are present in all income terciles but are largest among workers in the lowest prior-income groups, indicating that the returns to remote eligibility are not confined to already advantaged workers.
This pattern highlights a previously underappreciated distributional dimension of remote work, showing that its associated career gains extend well beyond the top of the earnings distribution.

\subsection*{Results}

\subsubsection*{Data and Identification Strategy}

To examine how remote-eligible job structures reshape career mobility, we combine two large-scale datasets that capture both realized job transitions and the supply of remote work opportunities. The first dataset, from Revelio Labs, contains longitudinal employment histories constructed from public professional profiles, allowing us to observe detailed job-entry events including wages, seniority levels, occupations, employers, and metropolitan locations. The second dataset, Cosmos, aggregates employer-posted vacancies and provides standardized information on job characteristics, including remote and hybrid work arrangements. 

By linking job-entry events to contemporaneous vacancy postings at the hiring organization, we construct a supply-side measure of remote eligibility at the point of hire. For each transition, we match the destination employer’s postings within a three-month pre-hire window and classify a job as remote-eligible when the majority of matched postings indicate remote or hybrid arrangements. This procedure assigns remote eligibility based on the opportunity structure facing entrants rather than individual negotiation outcomes. Our primary analytic sample includes 48 million job-entry events in the United States between 2020 and 2024, including approximately one million transitions into remote-eligible roles.

Because workers entering remote-eligible and fully on-site jobs may differ in their characteristics, we implement an inverse-probability-weighted regression framework to approximate comparable job transitions. On-site entries are reweighted to match remote-eligible transitions on observable pre-entry characteristics—including demographic attributes, prior wages, seniority, occupation, industry, previous employer characteristics, and job-entry-year indicators—and treatment effects are estimated within the resulting weighted sample while absorbing metropolitan-area fixed effects. Supplementary analyses using within-individual comparisons yield similar effect magnitudes, suggesting that the results are unlikely to be driven by stable individual differences (Supplementary Section S3). Additional technical details are described in the Statistical analysis section of the Methods.

\subsubsection*{Descriptive Evidence}
\begin{figure} 
	\centering
	\includegraphics[width=\textwidth]{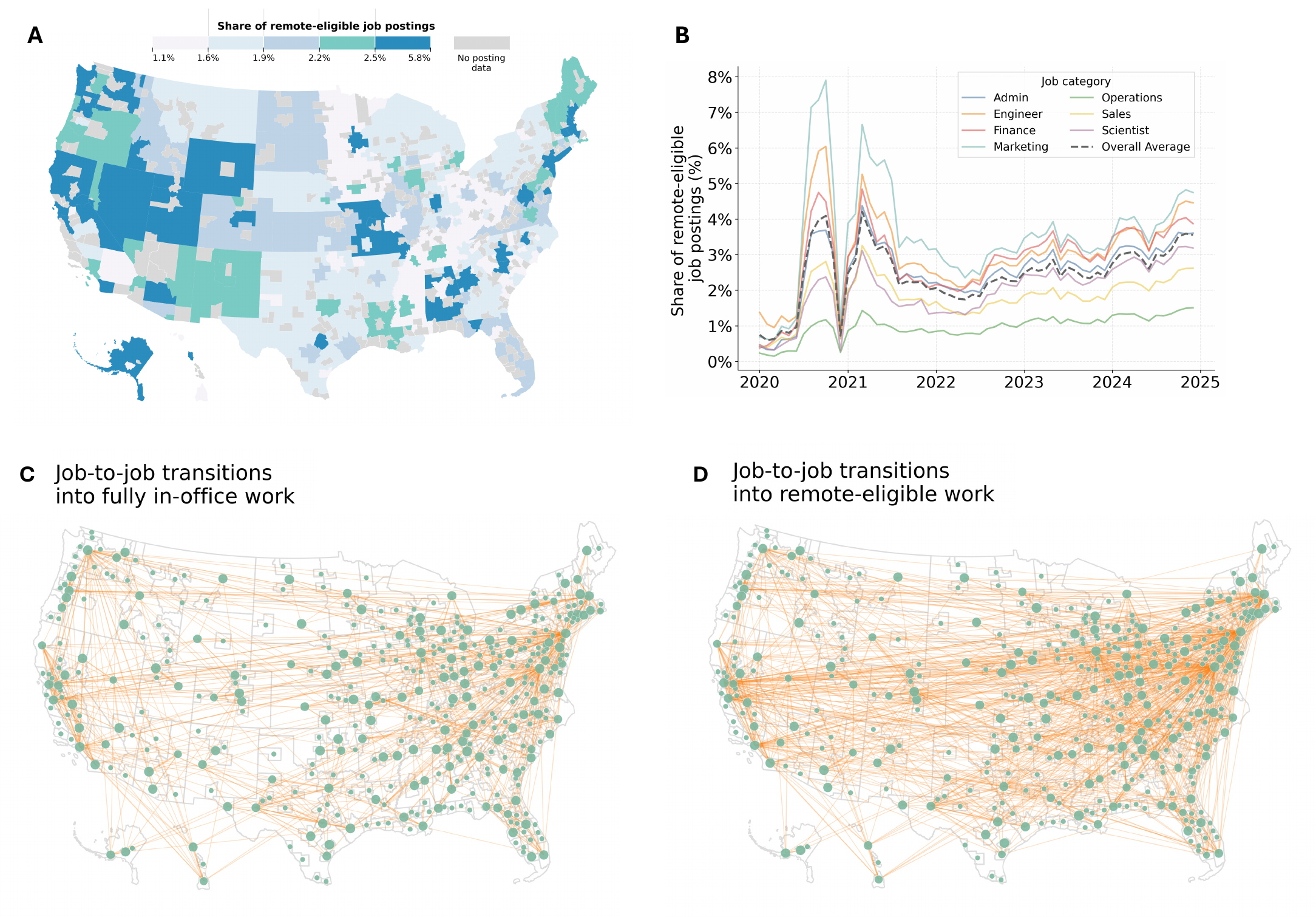} 

	\caption{\textbf{Remote-eligible jobs in the United States: job postings and job transitions.}
(\textbf{A}) Share of job postings classified as remote-eligible across U.S. metropolitan statistical areas (MSAs) and non-metropolitan counties, aggregated over 2020--2024.
(\textbf{B}) Monthly national time series of the share of remote-eligible job postings by major occupational category from 2020 to 2024.
(\textbf{C}) Spatial network of inter-metropolitan job transitions into fully in-office positions.
(\textbf{D}) Spatial network of inter-metropolitan job transitions into remote-eligible positions.
Nodes represent metropolitan and non-metropolitan areas, with node sizes indicating the within-area job retention share. 
Edges represent inter-metropolitan job transitions, with edge widths representing the share of job transitions from an origin area to a destination area. 
Edge weights and node sizes are defined based on transition shares and retention rates. Transitions into remote-eligible positions exhibit a denser pattern of cross-metro job transitions than transitions into fully in-office positions.}
	\label{figure_map} 
\end{figure}

In our data, more than 12 million remote-capable job opportunities were posted in the United States between 2020 and 2024 (roughly 2.4 million annually), accounting for about 2.3\% of all vacancies during the period and indicating the emergence of a persistent segment of the labor market. As shown in Fig.~\ref{figure_map}A, remote eligibility has spread across a wide geographic footprint by 2024, with notably high shares not only in coastal metropolitan areas but also across many interior Metropolitan Statistical Areas (MSAs). This pattern contrasts with the pre-pandemic concentration of flexible work opportunities in a small set of technology-oriented hubs \cite{dingel2020many,ozimek2020future}, indicating a broader structural shift in where remote-capable jobs are created.

Temporally, Fig.~\ref{figure_map}B shows that the rise of remote-eligible work has persisted well beyond the early pandemic shock. All occupations exhibited a sharp increase in remote eligibility during the onset of the COVID-19 pandemic, after which levels declined but settled at a new and substantially higher baseline compared with pre-pandemic conditions. The pattern is most pronounced in white-collar fields such as administration, marketing, engineering, and finance, which maintain higher remote shares than the overall average, while operations roles remain consistently low. Taken together, these trends indicate that flexible work arrangements have become a lasting element of U.S. labor demand.

Beyond changes in job postings, remote eligibility is also associated with more frequent inter-metropolitan job-to-job transitions. As illustrated in Fig.~\ref{figure_map}C–D, transitions into remote-eligible positions display a denser pattern of cross-metropolitan flows than transitions into fully in-office jobs. This pattern is consistent with remote eligibility expanding the geographic scope of feasible job matching across locations.
\FloatBarrier
\FloatBarrier

\begin{figure}[!htbp] 
	\centering
	\includegraphics[width=\textwidth]{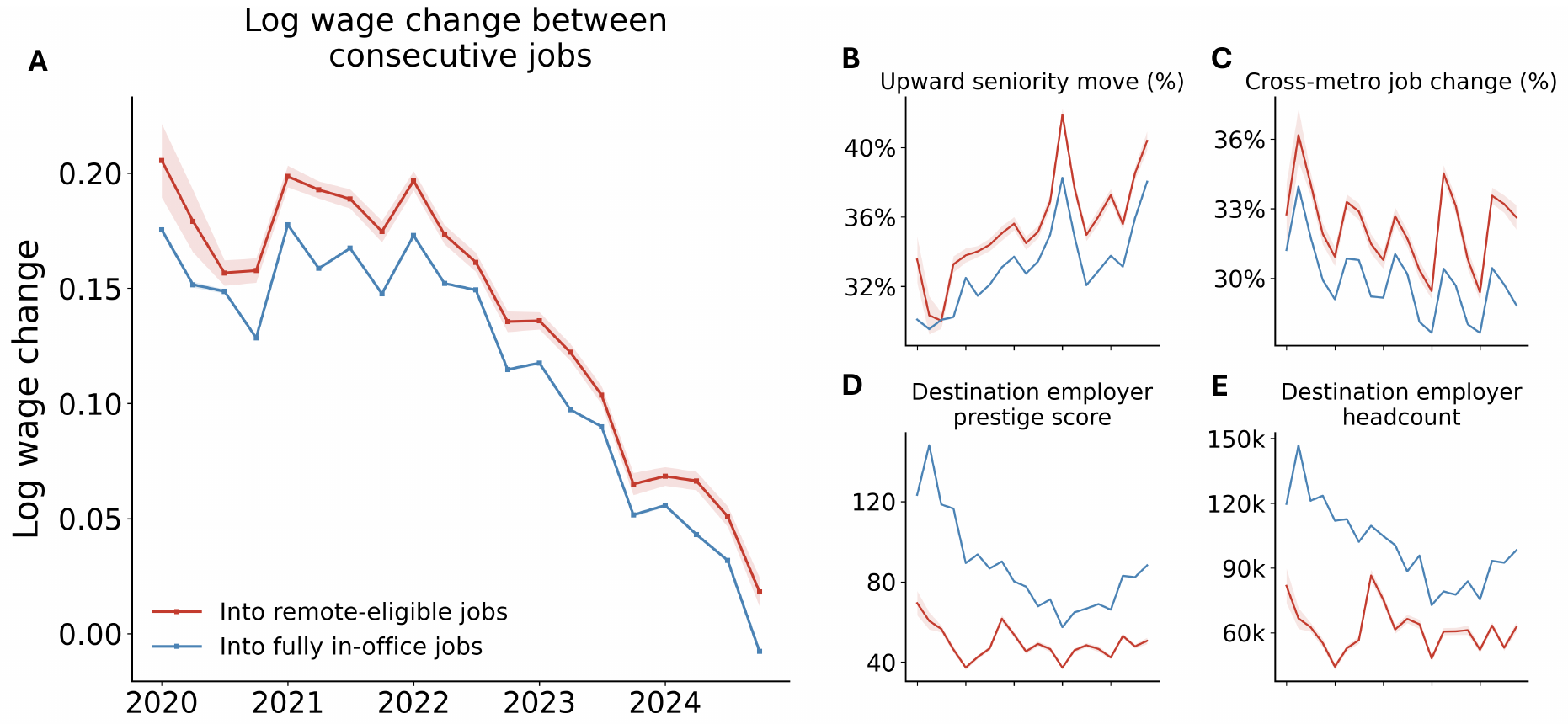} 

	\caption{\textbf{Comparisons between remote-eligible jobs and fully in-office jobs across career outcomes and job transition patterns.}
(\textbf{A}) Average log wage change between a worker’s previous job and destination job at consecutive job-to-job transitions into remote-eligible versus fully in-office positions. (\textbf{B}) Share of job transitions involving an upward seniority move, defined as an increase in seniority level relative to the worker’s previous position. (\textbf{C}) Share of job transitions involving a cross-metropolitan job change. (\textbf{D}) Average prestige score of destination employers.
(\textbf{E}) Average headcount of destination employers.
Panels (\textbf{A})--(\textbf{E}) report quarterly averages over the 2020--2024 period in the United States; red lines denote remote-eligible transitions and blue lines denote fully in-office transitions.
}
	\label{figure_main_descriptive} 
\end{figure}

Against this backdrop, we next examine how entry into remote-eligible jobs shapes individual career outcomes. Figure~\ref{figure_main_descriptive}A reveals a striking and persistent wage advantage associated with remote-eligible transitions: throughout the 2020–2024 period, workers entering remote-eligible positions experience substantially higher wage growth at job transitions than those entering fully in-office roles, with the gap remaining stable over time.

This wage premium is mirrored along other dimensions of career advancement. As shown in Fig.~\ref{figure_main_descriptive}B, remote-eligible entrants are more likely to experience upward seniority moves. At the same time, remote-eligible transitions are significantly more likely to involve cross-metropolitan job changes (Fig.~\ref{figure_main_descriptive}C), with the transition networks in Fig.~\ref{figure_map}C–D providing a visual counterpart to this pattern.

These gains are accompanied by systematic differences in destination employers: relative to in-office entrants, remote-eligible workers sort into employers with lower prestige scores and smaller headcounts (Fig.~\ref{figure_main_descriptive}D–E), suggesting that career advancement through remote-eligible transitions often occurs outside traditional pathways tied to large or prestigious employers.

\FloatBarrier
\FloatBarrier

\subsubsection*{Remote Work as a Driver of Wage and Career Mobility}

To estimate the effects of remote work eligibility on career outcomes, we combine inverse probability weighting with fixed-effects regression to address non-random sorting into remote jobs (see Methods for details). In contrast to the compensating-differentials framework that predicts wage penalties for work flexibility, the results reveal a statistically robust wage premium associated with entering a remote-eligible job.

\begin{figure} 
	\centering
	\includegraphics[width=\textwidth]{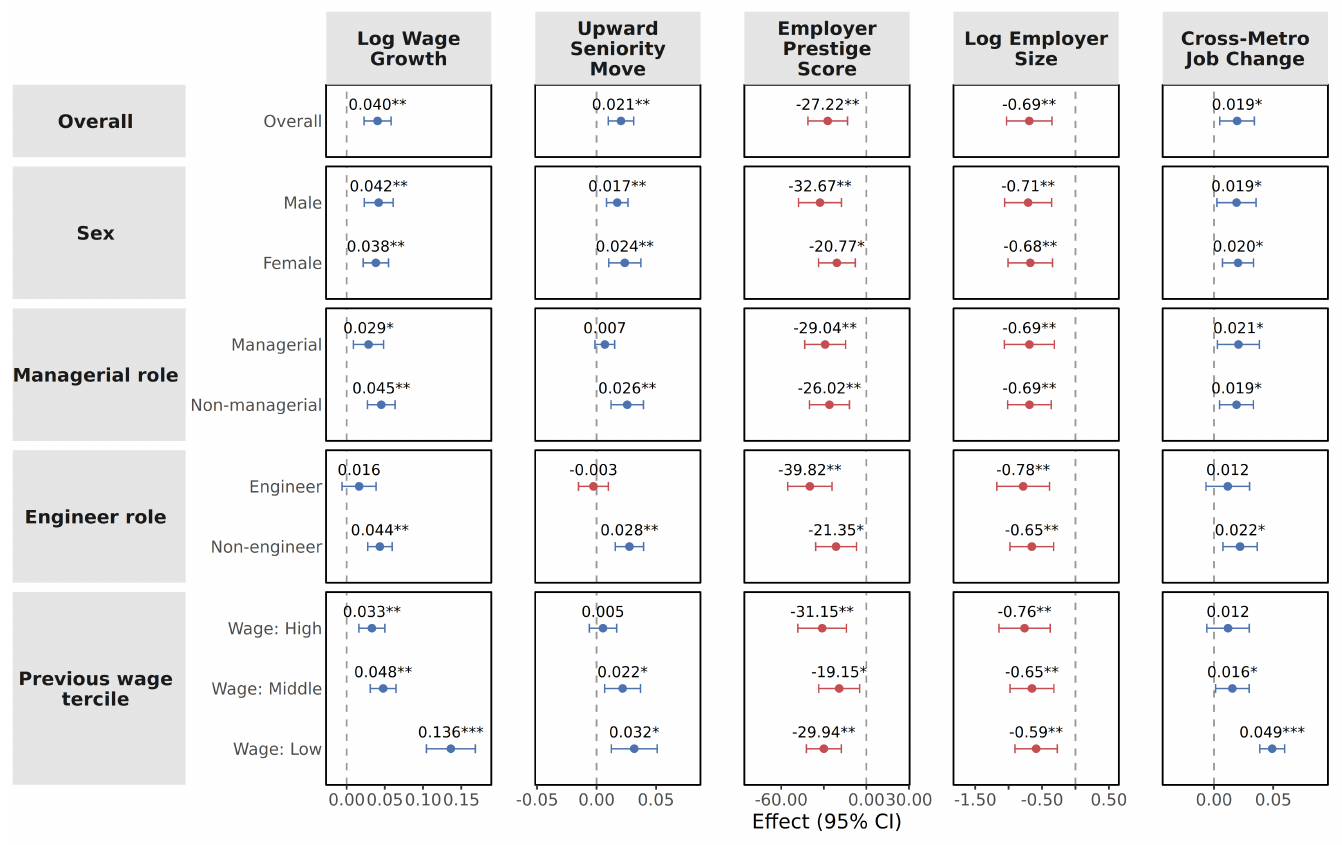} 

	\caption{\textbf{Effects of remote eligibility on individuals’ career outcomes.}
Estimated treatment effects of entering a remote-eligible job on five outcomes: log wage change between consecutive jobs, upward seniority move, prestige score of the destination employer, log headcount of the destination employer, and the probability of a cross-metropolitan job change. Estimates are reported for the full sample and across worker subgroups defined by sex, previous-job managerial status, previous-job engineering role, and terciles of previous wages. Points indicate coefficient estimates and horizontal bars denote 95\% confidence intervals. Statistical significance is indicated as $^{***}p<0.001$, $^{**}p<0.01$, $^{*}p<0.05$, and $^{.}p<0.1$.}
	\label{figure_main_effect} 
\end{figure}

Workers who transition into remote-eligible jobs experience systematically higher wage growth than comparable on-site entrants. As shown in Fig.~\ref{figure_main_effect}, this advantage amounts to a 0.04-log-point increase in wage growth, corresponding to an additional 4.13\% increase in earnings relative to workers’ previous jobs. This premium is robust across gender groups and across managerial and non-managerial roles. However, it is concentrated outside engineering occupations and is absent among engineers, suggesting that the observed effects are unlikely to be driven by a technology-sector boom concentrated in engineering roles. The gains are largest for workers in the lowest previous-wage tercile, indicating that remote eligibility disproportionately benefits individuals with more limited prior earning power.

Remote eligibility also increases movement into higher-ranked roles. Entrants to remote-eligible positions are 2.05 percentage points more likely to move into jobs with higher seniority than observationally similar workers entering fully in-office roles (Fig.~\ref{figure_main_effect}, column 2). This pattern indicates that remote eligibility is associated with upward occupational mobility, rather than offering flexibility at the expense of career progression. Because job titles may have different economic meanings across employers of different sizes, we assess the robustness of this result using alternative, employer-size-adjusted measures of upward seniority moves based on a calibrated wage--seniority space. The estimated effects on the probability and magnitude of upward seniority moves remain positive and statistically significant under these alternative definitions (Supplementary Section 5).
\FloatBarrier
\FloatBarrier

A striking organizational pattern accompanies these gains. Workers transitioning into remote-eligible jobs move toward employers with lower prestige and smaller organizational scale. As shown in Fig.~\ref{figure_main_effect} (columns 3), headcount at destination organizations declines by about 0.7 log points, corresponding to roughly a 50\% reduction in organizational size. Consistent with this scale effect, destination organizations also exhibit markedly lower prestige scores (Fig.~\ref{figure_main_effect}, column 4). These patterns indicate that remote work does not primarily channel workers toward large, established, high-status organizations. Instead, it facilitates movement into a broader and more diverse set of employers—often smaller and more agile organizations that use flexibility and title progression to compete for talent.

Remote eligibility is also associated with a pronounced expansion in the spatial scope of job changes. The probability of a cross-metro job change rises by approximately 1.95 percentage points for workers entering remote-eligible positions, indicating that remote-eligible jobs meaningfully relax location-based constraints in job matching. This mobility response is strongest among workers in the lowest previous-wage tercile, suggesting that remote eligibility most effectively expands geographic opportunity for workers with the most limited local labor market options.

To assess whether the estimated effects are concentrated in industries with historically higher adoption of remote work, we stratify workers by the remote-work intensity of their current-job industry. As shown in Supplementary Fig. S3, the positive effects of remote eligibility on wage growth and upward seniority moves are strongest in low– and medium–remote-intensity industries and are attenuated in high–remote-intensity industries. This pattern suggests that the estimated effects are not confined to sectors where remote work is already prevalent, but instead reflect larger marginal gains in segments of the labor market where remote opportunities remain scarce. We obtain qualitatively similar patterns when stratifying workers by the remote-work intensity of their previous-job industry (Supplementary Fig. S4).

\subsubsection*{Temporal Persistence and Spatial Heterogeneity of Remote Eligibility Effects}
\begin{figure} 
	\centering
	\includegraphics[width=\textwidth]{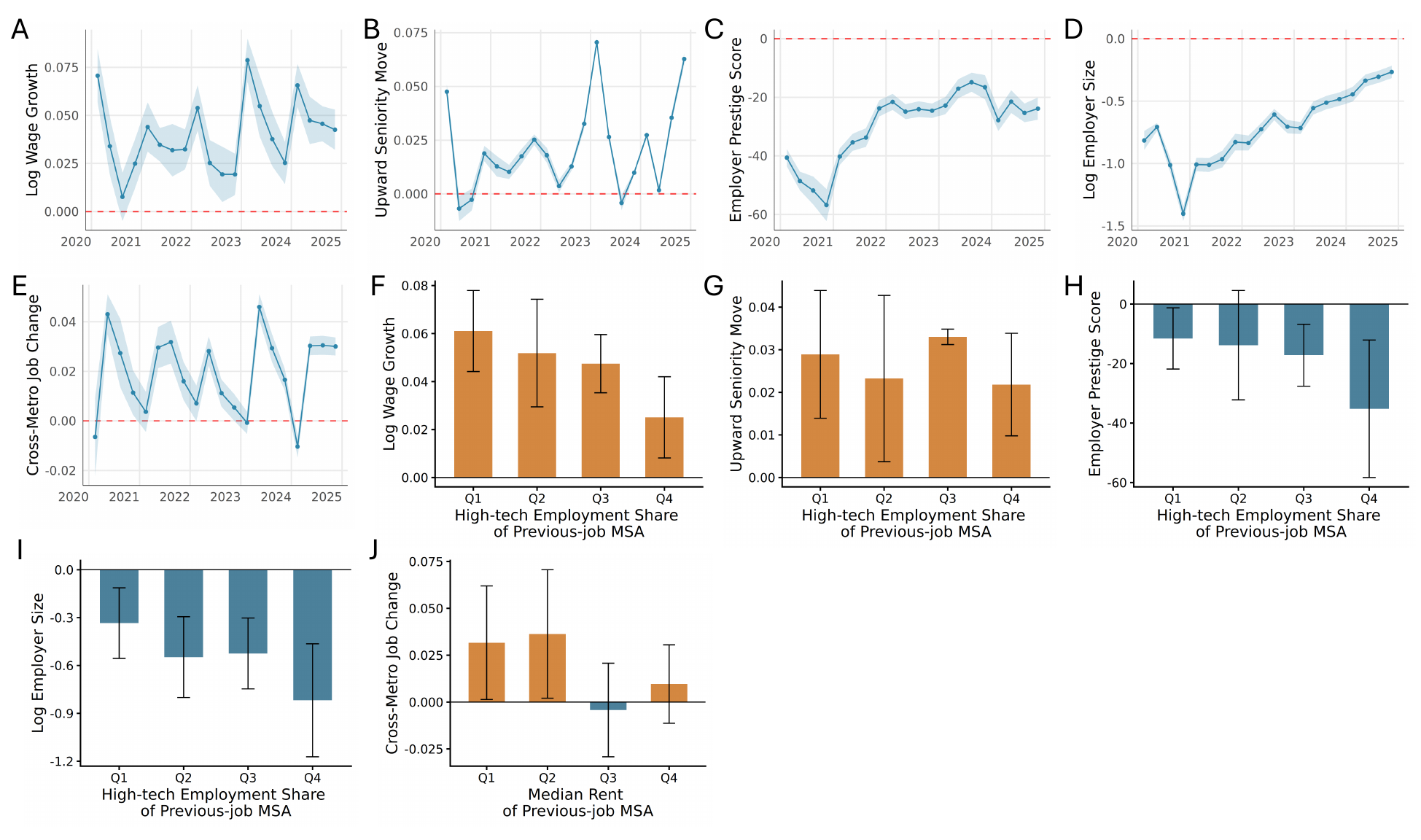} 
	\caption{\textbf{Temporal and regional heterogeneity of the remote-work effects.} 
	Panels A–E report estimates of the effects of entering remote-eligible jobs on five career outcomes by job-entry month: log wage growth (A), upward seniority move (B), employer prestige score (C), log employer size (D), and cross-metro job change (E). Points denote coefficient estimates by job-entry month, and shaded bands indicate 95\% confidence intervals. 
	Panels F–I summarize heterogeneity in average effects by the high-tech employment share of the worker’s previous-job metropolitan or non-metropolitan areas, shown in quartiles (Q1–Q4), for log wage growth (F), upward seniority move (G), employer prestige (H), and log employer size (I). 
	Panel J shows corresponding heterogeneity in cross-metro job changes by the median rent of the previous-job MSA. Bars indicate average effect estimates, and error bars indicate 95\% confidence intervals.
Across outcomes, remote eligibility is associated with larger gains in wage growth, higher probabilities of upward seniority moves, and more frequent cross-metro job transitions for workers originating from less advantaged regions, characterized by lower high-tech employment shares or lower housing costs. By contrast, remote-eligible transitions are, on average, directed toward smaller and less prestigious employers, with this pattern being more pronounced in regions with higher concentrations of high-tech employment.}
	\label{figure_hetero_effect} 
\end{figure}

A central question is whether the remote-work premium reflects a transient pandemic shock or a durable change in labor-market functioning. The temporal dynamics in Fig.~\ref{figure_hetero_effect} A–E point clearly toward persistence. The estimated effects on wage growth and the probability of upward seniority moves remain positive and statistically significant through 2024 (Fig.~\ref{figure_hetero_effect}A and B). The impact on cross-metro job changes (Fig.~\ref{figure_hetero_effect}E) exhibits seasonal cycles—likely mirroring broader hiring rhythms—but its baseline remains consistently above zero, with no evidence of structural attenuation. Together, these patterns suggest that remote work has introduced a stable channel through which job matching becomes partially decoupled from physical location. The organizational margin shows a more adaptive trajectory. Early in the sample, remote-eligible movers tend to join smaller and lower-prestige organizations, but these gaps narrow over time (Fig.~\ref{figure_hetero_effect}, C and D). 

Spatial stratification by the characteristics of the MSA associated with a worker’s previous job reveals substantial geographic heterogeneity. Fig.~\ref{figure_hetero_effect}F shows a pronounced negative gradient: the wage premium is largest for workers transitioning from MSAs with the lowest concentration of high-tech employment (Q1) and is substantially weaker for those from established tech hubs (Q4). This pattern suggests that workers originating from MSAs with lower baseline concentrations of high-tech employment experience disproportionately larger gains from remote eligibility, consistent with the idea that remote work relaxes geographic constraints and partially equalizes access to high-value opportunities across locations.

A parallel pattern emerges for cross-metro job transitions. As shown in Fig.~\ref{figure_hetero_effect}J, the effect of remote eligibility on cross-metro job transitions is largest for workers whose previous jobs were located in MSAs within the lower tiers (Q1 and Q2) of the rent distribution. This pattern is consistent with an expansion of job-matching opportunities across space, whereby workers in more affordable but opportunity-limited regions can access positions in higher-opportunity labor markets through remote work arrangements. Importantly, such cross-metro transitions may reflect remote-enabled employment without residential relocation, as well as traditional relocation following expanded job opportunities. In either case, remote eligibility broadens the feasible job-search frontier for workers in less advantaged places.
\FloatBarrier
\FloatBarrier
\subsubsection*{Remote Work Delivers the Largest Gains to the Most Constrained Workers}

As shown in Fig.~\ref{figure_main_effect}, wage gains from remote eligibility are largest among workers in the lowest tercile of previous earnings, who experience a 14.6\% increase in wages (a 0.136-log-point gain) relative to their previous jobs. This gain is more than three times as large as those observed for workers in the middle and top terciles. The resulting monotonic gradient suggests that remote eligibility disproportionately benefits workers with lower prior earnings, consistent with a narrowing of disparities in wage progression across the earnings distribution.

The distributional gradient aligns with differences in access to local labor-market opportunities. Workers in the lowest wage terciles are disproportionately concentrated in labor markets where high-skill employment opportunities are relatively scarce. Consistent with this interpretation, stratifying workers by the high-tech employment share of the metropolitan area associated with their previous job reveals a similar opportunity gradient. As shown in Fig.~\ref{figure_hetero_effect}F, the wage premium from remote eligibility is largest for workers originating in MSAs with the lowest concentration of high-tech employment and declines steadily as local opportunity density increases. This pattern indicates that the largest gains accrue not simply to lower earners per se, but to workers whose local labor markets offer limited advancement prospects.

Additional evidence comes from spatial mobility responses. As illustrated in Fig.~\ref{figure_hetero_effect}J, cross-metro job changes respond most strongly to remote eligibility in lower-rent labor markets, which typically coincide with thinner local job opportunity sets. Together, these patterns suggest that remote work disproportionately benefits workers facing tighter geographic opportunity constraints. Rather than uniformly raising wages, remote eligibility expands access to higher-quality career ladders for workers located in places where such opportunities are otherwise scarce.

Together, these results suggest a common mechanism: remote work disproportionately benefits workers whose career progression was previously limited by restricted access to non-local employers.


\subsubsection*{Distributional patterns and channels underlying wage growth effects of remote eligibility} 
\begin{figure} 
	\centering
	\includegraphics[width=\textwidth]{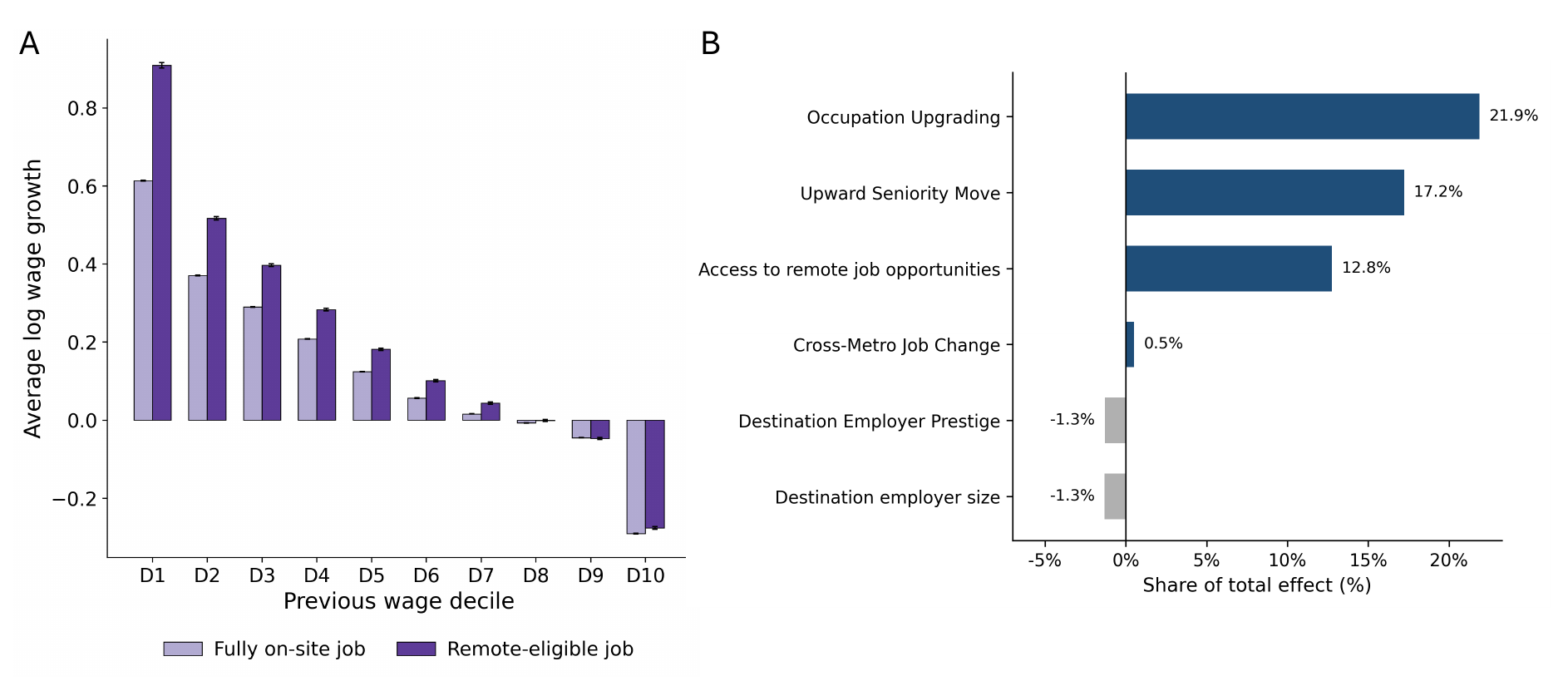} %
	\caption{\textbf{Wage growth differences by prior earnings and decomposition of the remote-work effect.} 
	(A) Average log wage growth following job entry, stratified by workers’ previous wage deciles (D1–D10), separately for fully on-site jobs and remote-eligible jobs. Wage growth exhibits a pronounced convergence pattern across the earnings distribution, with faster growth among workers in lower deciles and stagnation or decline among those in higher deciles. Across all deciles, entry into remote-eligible positions is associated with systematically higher wage growth relative to fully on-site entry. 
	(B) Decomposition of the total effect of remote eligibility on log wage growth into contributions from post-entry mechanisms, based on a Gelbach decomposition.
Bars report the share of the total effect attributable to each channel, including occupation upgrading, upward seniority move, access to remote job opportunities within a worker’s industry and metropolitan labor market, cross-metropolitan job change, employer prestige, and employer size. Positive values indicate channels that amplify wage growth, while negative values indicate offsetting contributions. The decomposition is conducted under the baseline identification framework and treats each channel as a post-transition mechanism, conditional on pre-transition characteristics.
}
	\label{figure_mechanisms} 
\end{figure}

To assess the distributional consequences and underlying channels through which remote work affects wage growth, we first examine wage growth stratified by workers’ previous earnings deciles (Fig.~\ref{figure_mechanisms}A). Wage growth displays a pronounced pattern of mean reversion: workers in the lowest deciles experience the fastest growth, whereas those in the highest deciles exhibit stagnation or decline. 

Against this backdrop, entry into remote-eligible positions is associated with systematically higher log wage growth than entry into fully on-site roles across the earnings distribution. This differential is largest in the bottom decile, where remote-eligible entrants experience substantially higher wage growth than their on-site counterparts. Even at the top of the distribution, where average wage growth turns negative, remote-eligible entry is associated with a smaller magnitude of wage decline relative to on-site transitions.

To unpack the sources of this differential under the baseline identification framework, Fig.~\ref{figure_mechanisms}B decomposes the total effect of remote eligibility on log wage growth into contributions from a set of post-transition job and labor-market channels using a Gelbach decomposition (see the Mechanism decomposition section of the Methods for details). In this decomposition, each channel is treated as a potential mechanism through which remote eligibility affects subsequent wage growth, conditional on pre-transition characteristics.

The decomposition results point to a common pattern across mechanisms: remote eligibility expands workers’ effective opportunity sets during job transitions, consistent with improved matching outcomes. Expanded access to remote job opportunities within a worker’s industry and metropolitan labor market accounts for 12.8\% of the total effect. This channel captures the facilitating role of a broader opportunity set, whereby remote eligibility increases the number and diversity of jobs workers can realistically consider when changing jobs.
\FloatBarrier
\FloatBarrier
Alongside expanded access to remote job opportunities, remote eligibility is associated with higher-quality job transitions along two closely related margins. Occupation upgrading accounts for the largest share of the estimated effect, explaining 21.9\% of the total impact, while upward seniority moves explain an additional 17.2\%. Taken together, these patterns are consistent with remote eligibility being associated with transitions into higher-paying and higher-rank positions. This interpretation is consistent with the idea that remote eligibility may relax geographic and organizational constraints that typically limit access to such positions, potentially reducing relocation, commuting, and coordination frictions faced by workers during job transitions.

Nevertheless, geographic relocation across metropolitan areas contributes little to overall wage growth: cross-metropolitan job changes explain only 0.5\% of the total effect. Organizational characteristics play a limited and partially offsetting role. Transitions toward smaller employers and lower-prestige organizations slightly reduce wage gains, together accounting for approximately -2.6\% of the total effect. This pattern suggests that the wage effects associated with remote eligibility are unlikely to be driven by compositional shifts toward smaller or less prestigious employers.

\subsection*{Discussion}
This study shows that eligibility for remote work reshapes career advancement by altering the structure of job transitions rather than merely offering short-term flexibility. In a labor market where remote work has settled into a stable and persistent mode of organization, our findings indicate that remote eligibility is associated with durable gains in wage growth, reflecting a combination of mechanisms during job transitions, including expanded effective opportunity sets, occupation upgrading, and higher rates of upward seniority moves. These gains arise alongside systematic shifts in organizational and spatial trajectories. Workers in remote-eligible roles are more likely to sort toward smaller and less prestigious organizations and to engage in cross-metropolitan job transitions, revealing a distinct pattern of upward mobility that does not rely on traditional career gateways tied to physical proximity or organizational status.

A critical implication of these results is that, in labor markets where remote work has become widespread—such as the United States in recent years—career advancement need not remain tightly coupled to physical proximity or organizational prestige. Traditional accounts of labor-market mobility emphasize agglomeration economies \cite{glaeser2010agglomeration,glaeser1999learning}, face-to-face learning \cite{storper2004buzz,battiston2021face}, and hierarchical sorting into large, high-status organizations \cite{idson1999workers,card2013workplace} as prerequisites for upward progression. In contrast, our evidence suggests that the expansion of remote-eligible work relaxes these constraints by decoupling access to employment opportunities from geographic proximity, allowing workers to compete for positions beyond their local labor markets without relocating. Workers who enter remote-eligible roles achieve faster wage growth and higher rates of upward seniority moves even as they sort away from established hubs and elite organizations. This pattern is difficult to reconcile with compensating-differentials models that predict wage penalties for flexibility and instead suggests that remote work is creating alternative pathways to upward mobility that are less tightly tied to traditional geographic and organizational hierarchies.

Importantly, the benefits of remote eligibility are not evenly distributed. We find that the gains are largest for workers in the lowest income terciles and for those originating from regions with relatively limited concentrations of high-tech employment. This pattern suggests that, although remote work delivers benefits across the income distribution, its gains are disproportionately concentrated among workers whose local labor markets offer fewer high-value employment opportunities.

These findings have two broader implications. First, perspectives that frame remote work primarily in terms of worker well-being, work–life balance, or short-run productivity overlook its deeper structural role in shaping career trajectories and labor-market efficiency. Our results show that remote eligibility operates through core mobility mechanisms—occupation upgrading, higher probabilities of upward seniority moves and expanded access to higher-quality job matches—rather than through compensating trade-offs. Second, ongoing efforts to re-anchor work strictly within physical offices risk underestimating the efficiency costs of constraining remote options. By expanding access to high-growth roles beyond a narrow set of employers and locations, remote eligibility acts as a lever for unlocking underutilized human capital, with the potential to improve allocative efficiency at the national scale and to mitigate long-standing regional disparities in economic opportunity.

We view this study as an initial step toward systematically documenting how remote work reshapes the structure of opportunity in modern labor markets. Several limitations are worth noting. 
First, while our identification strategy leverages within-individual variation and rich pre-entry controls, we cannot fully rule out time-varying confounding arising from occupational upgrading at the moment of job transition, whereby workers move into higher-paying occupations that are also more remote-feasible. However, several patterns in the data suggest that this mechanism is unlikely to be the primary driver of our findings: the estimated effects are not concentrated among high-wage, engineering, or traditionally remote occupations, nor among workers originating from high-tech metropolitan areas, and remain robust when restricting attention to within-individual, same-occupation switches or absorbing occupation-by-entry-year fixed effects (Supplementary Section 7).

Second, our analysis focuses on eligibility for remote work at the point of job entry, capturing how job structure shapes career transitions rather than the realized intensity or duration of remote work after hiring. As a result, our estimates reflect the career effects associated with access to remote-eligible positions rather than the direct impact of working remotely itself.
Third, while we identify occupation upgrading, upward seniority mobility and expanded opportunity sets as central mechanisms, we do not directly observe within-organization dynamics that may also be affected by remote work, such as mentorship, tacit knowledge transfer, or network formation, which could influence career trajectories over longer horizons.
Finally, our evidence is drawn from the post-2020 U.S. labor market, which features relatively high levels of remote work adoption and labor mobility. Whether similar mechanisms operate in other countries or institutional contexts remains an open question for future research.

\clearpage

\clearpage
\subsection*{Methods}

\subsubsection*{Data sources and coverage}
Our data are drawn from Revelio Labs, a workforce intelligence company that aggregates and standardizes labor market information from publicly available sources to construct a global workforce database \cite{revelio_data_dictionary_2025}. We use three integrated components: (1) individual-level career histories documenting job spells and job-to-job transitions; (2) organization-level job posting data capturing labor demand and work arrangement characteristics, including remote-work eligibility indicators; and (3) individual profile data providing demographic and background covariates. These components are obtained from Revelio Labs’ career history records, job posting data, and user profile data.

\paragraph*{Individual career histories.} 
Individual career history data are derived from publicly available professional profiles (for example, LinkedIn) curated and provided by Revelio Labs \cite{revelio_data_dictionary_2025}. These data record employment spells, including employer identifiers, job start and end dates, occupation codes, seniority levels, salaries, and geographic locations. Salaries are predicted at the position level using a regression-based model that incorporates occupation, seniority, employer, and country characteristics by Revelio Labs. The model is trained on over 200 million observed salary records from job postings and publicly available labor certification applications and achieves an out-of-sample root mean squared error (RMSE) of approximately 14\%. Location information is obtained from raw location fields reported in online profiles and mapped to metropolitan or non-metropolitan areas. Each record corresponds to one employment spell, enabling reconstruction of job-to-job transitions over time.

Our unit of analysis is a job-entry event, defined as the start of a new position following a previous employment spell for the same individual. Across all raw position records observed as of the end of 2024, the data comprise approximately 1.7 billion employment spells spanning 606 million unique individuals and 26 million organizations worldwide.

We focus on job-entry events in the United States between 2020 and 2024, yielding an analytical sample of approximately 85 million job-entry events involving 37 million unique individuals and 3.7 million organizations.

\paragraph*{Job posting data.} 
Job posting data are drawn from COSMOS, Revelio Labs’ unified and deduplicated job postings database \cite{revelio_cosmos_2025}. Each posting is associated with a unique employer identifier, an occupation category (position roles classified into 50 discrete levels), country and (non-)metropolitan area information, posting date, and a standardized classification of work arrangement (fully in office, partially in office, fully remote, territory remote, and temporary remote).

We link individual job-entry events to employer job posting data and aggregate postings at the \emph{employer $\times$ occupation $\times$ location $\times$ year-month} level, producing monthly counts of postings by work arrangement category. Focusing on the United States between 2020 and 2024, the aggregated dataset contains approximately 533 million job postings, collapsed into 95 million employer--occupation--location--month cells.

\paragraph*{User profile data.}
We link user profile data to individuals’ career history records to enrich job-entry observations with demographic attributes. Gender is inferred by Revelio Labs based on the likelihood of the first name being male or female, while ethnicity is estimated by Revelio Labs using first and last names in combination with individuals’ geographic location \cite{revelio_data_dictionary_2025}.

\subsubsection*{Construction of job-entry events and outcomes}

\paragraph*{Job-entry events and previous-job linkage.}
For each individual, job spells are ordered chronologically by start date. A job-entry event is defined as the start date of a new job spell. For each entry, we identify the immediately preceding job using lagged employment records and construct previous-job characteristics, including prior employer identifiers, occupation, seniority, wage, industry, metropolitan area, employer size, and employer prestige.

\paragraph*{Outcome variables.}
Let $i$ index job-entry observations. For each entry $i$, let $t(i)$ denote the entry year, and let $t(i)-1$ refer to the worker’s immediately preceding job. We analyze multiple outcomes measured at job entry. Wage growth is defined as the log change in salary between the entry job and the previous job:
\begin{equation}
\Delta \ln(\text{wage}_i) = \ln(\text{wage}_{i,t(i)}) - \ln(\text{wage}_{i,t(i)-1}).
\end{equation}

Upward seniority move is a binary indicator equal to one if a worker transitions to a job with a higher seniority level than in the previous job spell. Seniority is coded as a seven-level ordered categorical variable derived from standardized job-title classifications provided by Revelio Labs, spanning entry-level to senior executive positions:
\begin{equation}
\text{UpwardSeniorityMove}_i
=
\mathbb{1}\!\left\{
\text{Seniority}_{i,t} > \text{Seniority}_{i,t-1}
\right\}.
\end{equation}

Cross-metropolitan job change is defined as an indicator equal to one if a job entry involves a change in metropolitan or non-metropolitan area relative to the previous job:
\begin{equation}
\text{Move}_i = \mathbb{1}\{\text{Metro}_{i,t(i)} \neq \text{Metro}_{i,t(i)-1}\}.
\end{equation}

We also examine organizational outcomes at job entry, including parent-organization size and parent-organization prestige. Parent-organization size is measured as $\log(1+\text{Headcount})$, where Headcount denotes the annual employment count at the ultimate-parent organization level. We compute Headcount globally from observed employment spells by expanding each spell over the calendar years it spans and counting the number of unique individuals employed by each ultimate parent in each year. In the job-entry analysis, we use the parent-organization Headcount in the entry year.

Parent-organization prestige is constructed from a directed inter-organization labor-flow network based on global job-to-job transitions observed from 2020 onward. Nodes represent ultimate-parent organizations, and weighted edges capture the number of observed worker transitions between organizations. Prestige is measured using a standardized PageRank score computed on this network, with higher values indicating greater centrality in global labor flows. The labor-flow network is constructed using global transition data rather than being restricted to the U.S. sample in order to capture organizations’ positions in the broader international labor market. For reference, Table~S3 lists the top 20 U.S. employers by headcount, while Table~S4 reports the top 20 U.S. employers ranked by company prestige.

\subsubsection*{Measuring remote-work eligibility at job entry}

\paragraph*{Matching job entries to postings.}
To characterize the remote-work structure of jobs at the time of entry, each job-entry event is linked to employer job postings from the same employer, occupation category, country, and metropolitan area within a pre-entry window of $K=3$ months, defined as $[\text{EntryMonth}_i - K,\ \text{EntryMonth}_i]$. If multiple posting months fall within this window, the most recent month prior to entry is used to characterize the job’s remote-work eligibility. In robustness analyses, we vary the length of the pre-entry matching window to six months and obtain substantively similar results (Supplementary Table S8).

Applying this matching procedure yields an analytical sample of approximately 48 million job-entry events with defined remote-work eligibility, corresponding to 55.8\% of all job-entry events in the U.S. sample.

\paragraph*{Remote share definition.}
Let $f$, $o$, $l$, and $m$ index the employer, occupation, location, and calendar year--month, respectively. For each employer--occupation--location--month cell, we compute the share of postings offering any form of remote work:
\begin{equation}
\text{RemoteShare}_{f,o,l,m}
=
\frac{
\text{FullyRemote}
+ \text{TerritoryRemote}
+ \text{TemporaryRemote}
+ \text{PartialOffice}
}{
\text{TotalPostings}
}.
\end{equation}

\paragraph*{Treatment definition.}
We define the treatment variable as a binary indicator of remote eligibility at job entry. A job entry is classified as \emph{remote-eligible} if the share of remote postings in the matched employer--occupation--location--month cell exceeds 0.5:
\begin{equation}
\text{RemoteEligible}_i = \mathbb{1}\{\text{RemoteShare}_{f,o,l,m} \geq 0.5\}.
\end{equation}
Job entries with $\text{RemoteShare}_{f,o,l,m} < 0.5$ are classified as non-remote-eligible and serve as the control group. Results are robust to alternative threshold choices: using a more stringent cutoff of
0.8 to define remote eligibility yields qualitatively similar estimates (Supplementary Table S8).

\subsubsection*{Covariates}
All analyses adjust for a rich set of pre-entry covariates capturing demographic characteristics, prior earnings and career stage, occupation, previous employer characteristics, industry affiliation, geographic location, and entry-year effects. These covariates account for observable differences between workers entering remote-eligible and non-remote-eligible positions, following established literature on labor market sorting, compensating differentials, and the determinants of remote work adoption \cite{dingel2020many,mas2017valuing}.

Demographic characteristics include gender and ethnicity, measured using probabilistic attributes provided by Revelio Labs. Gender is inferred from first names using models informed by U.S. Social Security Administration data, while ethnicity is inferred from first and last names and geographic location using U.S. Census data \cite{revelio_data_dictionary_2025}. We further control for previous salary (log-transformed) and previous seniority to account for differences in pre-entry earnings capacity and career stage. 

Occupation fixed effects are included based on K50 role classifications, a proprietary occupation taxonomy that groups job titles into approximately 50 role categories. These controls capture systematic differences across occupations in remote feasibility, wage growth trajectories, and upward seniority mobility. 

Employer-level controls capture characteristics of the worker’s previous organization, including pay-setting practices and career progression dynamics \cite{card2013workplace,autor2020fall}. These controls include a PageRank-based measure of employer prestige constructed from inter-organization worker flows, parent-level employer headcount (log-transformed) as a proxy for employer size, and two-digit NAICS industry indicators.

Geographic and temporal controls are included to isolate the effect of remote eligibility from geographic wage premia and agglomeration economies. Specifically, we control for the metropolitan area of the previous job alongside entry-year fixed effects that absorb time-varying labor market conditions at the point of job entry \cite{glaeser2011cities,moretti2010local}.

\subsubsection*{Statistical analysis}

To quantify the effects of remote-work eligibility on career outcomes, we employ a weighting-based estimation strategy combined with fixed-effects regression. Because workers entering remote-eligible and fully on-site jobs may differ in their observable characteristics, we use inverse probability weighting (IPW) to approximate comparable job transitions.

Specifically, non-remote-eligible job-entry events are reweighted to match the distribution of observed pre-entry covariates among remote-eligible entrants, thereby reducing differences in observable characteristics between treated and control groups. The estimation proceeds in the following steps.

\paragraph*{Main estimation strategy.}
We begin by estimating propensity scores that estimate the probability of entering a remote-eligible position conditional on pre-entry characteristics. Propensity scores are estimated using logistic regression models that relate remote eligibility at job entry to pre-entry covariates described above, including gender and ethnicity, previous-job salary and seniority, previous-job occupation and industry, previous employer characteristics (prestige and size), and the job-entry year.

The estimated propensity scores are then used to construct inverse probability weights targeting the average treatment effect on the treated (ATT). Remote-eligible job-entry events receive a weight of one, while non-remote-eligible events are weighted by the estimated odds of treatment ($\frac{p_i}{1-p_i}$) based on their propensity scores. Covariate balance after weighting is assessed using standardized mean differences (SMDs) for each covariate included in the propensity score model (see Supplementary Section 2). Across all main specifications, the maximum absolute SMD after weighting is below 0.1, indicating satisfactory balance.

We then estimate treatment effects using weighted fixed-effects regression models of the following form:

\begin{equation}
Y_i = \beta\,\text{RemoteEligible}_i + Z_i^\top \gamma +
\mu_{s(i)} + \mu_{e(i)} + \mu_{o(i)} + \mu_{n(i)} + \mu_{m(i)} + \mu_{t(i)} + \varepsilon_i,
\label{equ:main}
\end{equation}
where $Y_i$ denotes the outcome associated with job-entry event $i$. The vector $Z_i$ includes continuous pre-entry controls, namely previous wage (log-transformed), previous employer structural prestige, and previous employer size (parent-level headcount; log-transformed), as well as the predicted probability of being female. 

The fixed effects $\mu_{s(i)}$, $\mu_{e(i)}$, $\mu_{o(i)}$, $\mu_{n(i)}$, $\mu_{m(i)}$, and $\mu_{t(i)}$ denote fixed effects for previous seniority level, ethnicity, occupation, industry, metropolitan area, and entry year, respectively. Metropolitan-area fixed effects are included in the regression stage to account for geographic wage premia and local labor-market conditions.

All regressions are weighted by the ATT weights defined above. Standard errors are two-way clustered by previous metropolitan area and entry year to account for spatial and temporal dependence.

\paragraph*{Subgroup outcome analyses.}
To examine heterogeneity in the estimated effects, we report subgroup-specific estimates by re-estimating the outcome models within subsamples defined by worker and job characteristics. We examine heterogeneity by gender and prior wage terciles to assess whether the estimated effects are broadly distributed across workers or concentrated among those with greater pre-existing labor-market advantages. We also estimate effects separately by managerial status (managerial versus non-managerial roles) and by role type (engineering versus non-engineering occupations). These distinctions help assess whether the estimated effects vary across segments of the labor market characterized by differences in job autonomy and technological intensity. \cite{gibbs2023work,mas2017valuing}.

\paragraph*{Within-individual matched-pairs robustness analysis.}
To probe robustness to unobserved, time-invariant worker heterogeneity, we implement a within-individual matched-pairs design. We retain individuals who experience at least one remote-eligible job entry and at least one non-remote job entry during 2020--2024. For each remote-eligible job-entry event, we identify the temporally closest non-remote job-entry event for the same individual, forming matched pairs that isolate within-individual variation in remote eligibility.

Using the matched-pairs sample, we estimate pair fixed-effects regression models of the following form:
\begin{equation}
Y_{ip} = \beta \,\text{RemoteEligible}_{ip} + \lambda_{t(ip)} + \alpha_p + \varepsilon_{ip},
\end{equation}
where $Y_{ip}$ denotes the outcome associated with job-entry observation $i$ in matched pair $p$, $\alpha_p$ denotes matched-pair fixed effects, and $\lambda_{t(ip)}$ captures entry-year fixed effects. Standard errors are clustered at the matched-pair level. The final matched-pairs dataset consists of 515{,}034 matched pairs (1{,}030{,}068 job-entry observations), corresponding to 487{,}301 unique individuals. Results are reported in Supplementary Table~S8.

\subsubsection*{Mechanism decomposition}
To understand how remote eligibility relates to higher wage growth at job entry, we apply a Gelbach decomposition \cite{gelbach2016covariates} to apportion the estimated remote wage premium across observable post-entry channels. We consider six candidate mechanisms: upward seniority move, cross-metropolitan job change, subsequent employer size, subsequent employer prestige, the availability of remote job opportunities in the destination labour market, and occupation upgrading. Remote job availability is constructed at the metropolitan area--industry--year level as the share of job postings classified as remote-eligible.

Occupation upgrading is measured using changes in the relative wage position of workers’ occupations at entry. Let $i$ index a job-entry observation and let $t(i)$ denote its entry year. For each entry year $t$, we compute the mean entry wage for each K50 occupation $o$, denoted $\overline{w}_{o,t}$, and define $\mathrm{rank}_t(o)$ as the within-year rank of occupations based on $\overline{w}_{o,t}$. We define occupation upgrading as the change in occupation wage rank between the entry occupation and previous occupation, both evaluated using the same entry-year ranking:
\[
\Delta \mathrm{OccWageRank}_i
= \mathrm{rank}_{t(i)}(o^{\text{entry}}_i)
- \mathrm{rank}_{t(i)}(o^{\text{prev}}_i).
\]

To implement the decomposition, we compare a baseline specification with an augmented specification that additionally includes the vector of mechanism variables $\mathbf{M}_i$:
\begin{equation}
Y_i = \tau\,\text{RemoteEligible}_i + \mathbf{M}_i^\top \delta + \mathbf{W}_i^\top \lambda + \varepsilon_i ,
\end{equation}
where $Y_i$ denotes log wage growth at job entry, $\mathbf{M}_i$ denotes the vector of mechanism variables described above, and $\mathbf{W}_i$ collects all pre-entry covariates and fixed effects included in the main specification. The baseline model is identical but omits $\mathbf{M}_i$.

The reduction in the estimated coefficient on remote eligibility between the baseline and augmented specifications captures the portion of the wage premium associated with the included mechanisms. The Gelbach decomposition allocates this reduction across components of $\mathbf{M}_i$ conditional on the others. For presentation, we express each mechanism’s contribution as a share of the total remote-eligibility effect on wage growth. All decomposition estimates are weighted using the inverse probability weights described above to account for selection into remote eligibility.


\clearpage 

%
\bibliography{bibliography} 
\bibliographystyle{sciencemag}

%
%
%
%
%
%




\paragraph*{Competing interests:}
There are no competing interests to declare.

\paragraph*{Data and materials availability:}
The data necessary to reproduce the figures and tables reported in this study can be accessed through Revelio Labs, subject to approval and the terms of a data-use agreement \cite{RevelioLabsResearch2025}.

The code used for analysis will be made publicly available upon publication. 
\clearpage
\appendix

\renewcommand{\thepage}{S\arabic{page}} 
\renewcommand{\thesection}{S\arabic{section}} 
\renewcommand{\thetable}{S\arabic{table}} 
\renewcommand{\thefigure}{S\arabic{figure}} 

\section*{Supplementary Information}

\tableofcontents

\vfill
\begin{center}
Note: Supplementary numbering follows the original submission version. For clarity and brevity, only selected supplementary materials are included in this preprint.
\end{center}

\newpage
\section{Data description}

This study combines three main datasets from Revelio Labs: individual career histories, job posting data, and user profile attributes. Career history data provide longitudinal employment spells for individual workers, including employer identifiers, occupation roles, start and end dates, seniority levels, and predicted wages. Job posting data describe employer hiring activity and job characteristics, including occupation classifications, geographic locations, and remote-work arrangements. User profile data provide demographic attributes such as gender probability and estimated ethnicity.

Across all records observed as of the end of 2024, the underlying career-history dataset contains approximately 1.7 billion employment spells covering 606 million unique individuals and 26 million organizations worldwide. Restricting attention to job-entry events in the United States between 2020 and 2024 yields approximately 85 million job-entry events involving 37 million individuals and 3.7 million organizations.

To identify remote-work eligibility at job entry, we link each job-entry event to employer job postings using a matching procedure based on employer identity, occupation classification, geographic location, and time. Applying this procedure yields approximately 48 million job-entry events with defined remote-work eligibility, corresponding to 55.8\% of all U.S. job-entry events during the study period.

\begin{figure}[!htbp] 
	\centering
	\includegraphics[width=\textwidth]{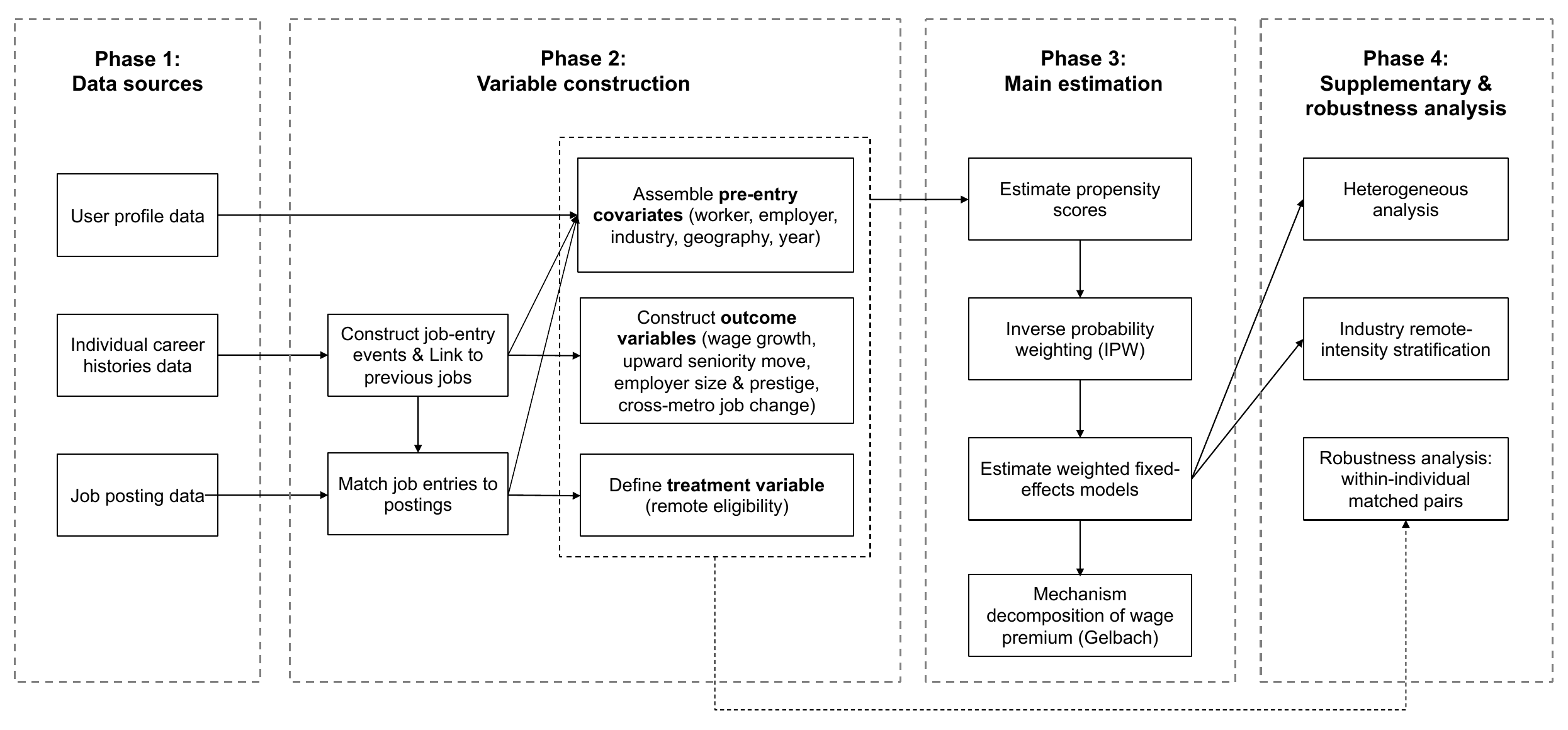} 

	\caption{\textbf{Data and analysis pipeline.}}
	\label{fig:pipeline} 
\end{figure}

The final analytical dataset used in the regression analysis is constructed by further requiring non-missing outcome and covariate information for the relevant estimation samples. Figure~\ref{fig:pipeline} summarizes the overall data integration and analysis pipeline.

\FloatBarrier
\FloatBarrier

\section{Covariate balance after inverse probability weighting}

To assess whether the inverse probability weighting procedure successfully balances observable characteristics between remote-eligible and non-remote-eligible job-entry events, we examine standardized mean differences (SMDs) for each covariate included in the propensity score model.

Tables~\ref{tab:tab:baseline_unweighted} and \ref{tab:tab:baseline_weighted} report baseline covariate balance before and after applying ATT weights. Continuous variables are summarized using means and standard deviations, and categorical variables using counts and percentages. The SMD for each covariate is defined as the difference in means between treated and control groups divided by the pooled standard deviation.

Before weighting, several covariates exhibit notable imbalance, particularly previous wages, employer characteristics, and industry composition. After applying inverse probability weights, covariate balance improves substantially across all variables. In the weighted sample, absolute SMDs for all covariates fall well below the commonly used threshold of 0.1, indicating satisfactory balance between treated and control job-entry events.

These results suggest that the weighting procedure effectively aligns the distributions of observed worker, employer, and job characteristics between remote-eligible and non-remote-eligible observations, supporting the validity of the weighted comparisons used in the main analysis.
\setcounter{table}{5}
\FloatBarrier

{\footnotesize

\begin{longtable}[t]{lrrr}
\caption{\label{tab:tab:baseline_unweighted}\textbf{Baseline characteristics and covariate balance (Unweighted)}. Continuous variables are reported as mean (SD); categorical variables as count (\%). Absolute standardized mean differences are shown for each row. The column-header N reports the raw number of observations. For presentation clarity, we report a compact set of previous-role (K50) levels: the 12 most frequent roles and the 8 roles with the largest unweighted absolute standardized differences (union of the two sets). }\\
\toprule
Baseline characteristics & \multicolumn{1}{c}{Treated}  & \multicolumn{1}{c}{Control} & \multicolumn{1}{c}{Absolute standardized}\\
 & \multicolumn{1}{c}{(N=512664)} & \multicolumn{1}{c}{(N=18923809)} & \multicolumn{1}{c}{mean difference}\\

\midrule
Female (prob.) & 0.46 (0.46) & 0.46 (0.45) & 0.0043\\
Log(1 + previous wage) & 11.10 (0.71) & 10.79 (0.85) & 0.4312\\
Previous employer prestige & 59.65 (175.76) & 82.30 (196.20) & 0.1289\\
Log(1 + previous employer headcount) & 7.43 (3.70) & 7.98 (3.78) & 0.1498\\
Ethnicity, no. (\%) &  &  & \\
\hspace{3mm}API & 83770 (16.34\%) & 4321818 (22.84\%) & 0.1757\\
\hspace{3mm}Black & 37363 (7.29\%) & 1204876 (6.37\%) & 0.0354\\
\hspace{3mm}Hispanic & 36677 (7.15\%) & 1555241 (8.22\%) & 0.0413\\
\hspace{3mm}Multiple & 1797 (0.35\%) & 173086 (0.91\%) & 0.0955\\
\hspace{3mm}Native & 310 (0.06\%) & 10604 (0.06\%) & 0.0018\\
\hspace{3mm}White & 352747 (68.81\%) & 11658184 (61.61\%) & 0.1554\\
Previous role (K50), no. (\%) &  &  & \\
\hspace{3mm}Software Engineer & 51420 (10.03\%) & 2017712 (10.66\%) & 0.0210\\
\hspace{3mm}Coordinator & 37286 (7.27\%) & 1515593 (8.01\%) & 0.0283\\
\hspace{3mm}Medical Rep & 20535 (4.01\%) & 1173430 (6.20\%) & 0.1120\\
\hspace{3mm}Mechanical Engineer & 17591 (3.43\%) & 772121 (4.08\%) & 0.0356\\
\hspace{3mm}Cashier & 11065 (2.16\%) & 763507 (4.03\%) & 0.1291\\
\hspace{3mm}Crew Member & 11580 (2.26\%) & 742221 (3.92\%) & 0.1119\\
\hspace{3mm}Sales Associate & 27816 (5.43\%) & 680949 (3.60\%) & 0.0807\\
\hspace{3mm}Accountant & 19242 (3.75\%) & 686713 (3.63\%) & 0.0066\\
\hspace{3mm}Scientist & 12605 (2.46\%) & 551624 (2.91\%) & 0.0295\\
\hspace{3mm}Financial Advisor & 14679 (2.86\%) & 541213 (2.86\%) & 0.0002\\
\hspace{3mm}Producer & 21207 (4.14\%) & 523952 (2.77\%) & 0.0687\\
\hspace{3mm}Data Analyst & 12429 (2.42\%) & 463271 (2.45\%) & 0.0015\\
\hspace{3mm}IT Specialist & 21288 (4.15\%) & 435086 (2.30\%) & 0.0929\\
\hspace{3mm}Distribution Specialist & 5443 (1.06\%) & 385228 (2.04\%) & 0.0950\\
\hspace{3mm}Designer & 13879 (2.71\%) & 303403 (1.60\%) & 0.0680\\
Previous seniority, no. (\%) &  &  & \\
\hspace{3mm}1 Entry & 144339 (28.15\%) & 6338092 (33.49\%) & 0.1187\\
\hspace{3mm}2 Junior & 157179 (30.66\%) & 5947496 (31.43\%) & 0.0167\\
\hspace{3mm}3 Associate & 65857 (12.85\%) & 2285277 (12.08\%) & 0.0230\\
\hspace{3mm}4 Manager & 65624 (12.80\%) & 2134202 (11.28\%) & 0.0456\\
\hspace{3mm}5 Director & 61335 (11.96\%) & 1775365 (9.38\%) & 0.0796\\
\hspace{3mm}6 Executive & 15720 (3.07\%) & 384802 (2.03\%) & 0.0599\\
\hspace{3mm}7 SrExecutive & 2610 (0.51\%) & 58575 (0.31\%) & 0.0280\\
Previous industry (NAICS-2), no. (\%) &  &  & \\
\hspace{3mm}11 & 741 (0.14\%) & 36150 (0.19\%) & 0.0122\\
\hspace{3mm}21 & 2661 (0.52\%) & 144560 (0.76\%) & 0.0341\\
\hspace{3mm}22 & 4514 (0.88\%) & 159236 (0.84\%) & 0.0042\\
\hspace{3mm}23 & 6642 (1.30\%) & 287063 (1.52\%) & 0.0196\\
\hspace{3mm}31 & 7328 (1.43\%) & 424101 (2.24\%) & 0.0684\\
\hspace{3mm}32 & 15210 (2.97\%) & 673852 (3.56\%) & 0.0350\\
\hspace{3mm}33 & 39840 (7.77\%) & 1523294 (8.05\%) & 0.0104\\
\hspace{3mm}42 & 6953 (1.36\%) & 291007 (1.54\%) & 0.0157\\
\hspace{3mm}44 & 7578 (1.48\%) & 455330 (2.41\%) & 0.0769\\
\hspace{3mm}45 & 14603 (2.85\%) & 893413 (4.72\%) & 0.1126\\
\hspace{3mm}48 & 9821 (1.92\%) & 379535 (2.01\%) & 0.0066\\
\hspace{3mm}49 & 1702 (0.33\%) & 77447 (0.41\%) & 0.0134\\
\hspace{3mm}51 & 95705 (18.67\%) & 2535456 (13.40\%) & 0.1352\\
\hspace{3mm}52 & 63837 (12.45\%) & 2111341 (11.16\%) & 0.0392\\
\hspace{3mm}53 & 6599 (1.29\%) & 250562 (1.32\%) & 0.0033\\
\hspace{3mm}54 & 86285 (16.83\%) & 3161736 (16.71\%) & 0.0033\\
\hspace{3mm}55 & 339 (0.07\%) & 12212 (0.06\%) & 0.0006\\
\hspace{3mm}56 & 15659 (3.05\%) & 586919 (3.10\%) & 0.0027\\
\hspace{3mm}61 & 48196 (9.40\%) & 1825051 (9.64\%) & 0.0083\\
\hspace{3mm}62 & 22778 (4.44\%) & 1189736 (6.29\%) & 0.0895\\
\hspace{3mm}71 & 5937 (1.16\%) & 229400 (1.21\%) & 0.0051\\
\hspace{3mm}72 & 6966 (1.36\%) & 461243 (2.44\%) & 0.0932\\
\hspace{3mm}81 & 15113 (2.95\%) & 425290 (2.25\%) & 0.0414\\
\hspace{3mm}92 & 27559 (5.38\%) & 783648 (4.14\%) & 0.0547\\
\hspace{3mm}99 & 98 (0.02\%) & 6227 (0.03\%) & 0.0100\\
Entry year, no. (\%) &  &  & \\
\hspace{3mm}2020 & 41226 (8.04\%) & 1818694 (9.61\%) & 0.0577\\
\hspace{3mm}2021 & 117771 (22.97\%) & 4679107 (24.73\%) & 0.0417\\
\hspace{3mm}2022 & 113299 (22.10\%) & 5498647 (29.06\%) & 0.1677\\
\hspace{3mm}2023 & 125921 (24.56\%) & 4184096 (22.11\%) & 0.0570\\
\hspace{3mm}2024 & 114447 (22.32\%) & 2743265 (14.50\%) & 0.1880\\
\bottomrule
\end{longtable}
}
\FloatBarrier

\FloatBarrier

{\footnotesize

\begin{longtable}[t]{l r r r}
\caption{\label{tab:tab:baseline_weighted}\textbf{Baseline characteristics and covariate balance (IPW-weighted).} All treated and control summary statistics (means/SDs for continuous variables and counts/percentages for categorical variables) are computed using IPW weights. The column-header N reports raw sample sizes.}\\
\toprule
Baseline characteristics & \multicolumn{1}{c}{Treated}  & \multicolumn{1}{c}{Control} & \multicolumn{1}{c}{Absolute standardized}\\
 & \multicolumn{1}{c}{(N=512664)} & \multicolumn{1}{c}{(N=18923809)} & \multicolumn{1}{c}{mean difference}\\

\midrule
Female (prob.) & 0.46 (0.46) & 0.46 (0.45) & 0.0003\\
Log(1 + previous wage) & 11.10 (0.71) & 11.10 (0.73) & 0.0001\\
Previous employer prestige & 59.65 (175.76) & 59.84 (154.48) & 0.0011\\
Log(1 + previous employer headcount) & 7.43 (3.70) & 7.43 (3.81) & 0.0001\\
Ethnicity, no. (\%) &  &  & \\
\hspace{3mm}API & 83770 (16.34\%) & 83774 (16.34\%) & 0.0000\\
\hspace{3mm}Black & 37363 (7.29\%) & 37385 (7.29\%) & 0.0002\\
\hspace{3mm}Hispanic & 36677 (7.15\%) & 36623 (7.14\%) & 0.0004\\
\hspace{3mm}Multiple & 1797 (0.35\%) & 1791 (0.35\%) & 0.0002\\
\hspace{3mm}Native & 310 (0.06\%) & 311 (0.06\%) & 0.0000\\
\hspace{3mm}White & 352747 (68.81\%) & 352749 (68.81\%) & 0.0001\\
Previous role (K50), no. (\%) &  &  & \\
\hspace{3mm}Software Engineer & 51420 (10.03\%) & 51308 (10.01\%) & 0.0007\\
\hspace{3mm}Coordinator & 37286 (7.27\%) & 37363 (7.29\%) & 0.0006\\
\hspace{3mm}Medical Rep & 20535 (4.01\%) & 20554 (4.01\%) & 0.0002\\
\hspace{3mm}Mechanical Engineer & 17591 (3.43\%) & 17589 (3.43\%) & 0.0000\\
\hspace{3mm}Cashier & 11065 (2.16\%) & 11077 (2.16\%) & 0.0002\\
\hspace{3mm}Crew Member & 11580 (2.26\%) & 11590 (2.26\%) & 0.0001\\
\hspace{3mm}Sales Associate & 27816 (5.43\%) & 27751 (5.41\%) & 0.0005\\
\hspace{3mm}Accountant & 19242 (3.75\%) & 19219 (3.75\%) & 0.0002\\
\hspace{3mm}Scientist & 12605 (2.46\%) & 12614 (2.46\%) & 0.0001\\
\hspace{3mm}Financial Advisor & 14679 (2.86\%) & 14692 (2.87\%) & 0.0002\\
\hspace{3mm}Producer & 21207 (4.14\%) & 21239 (4.14\%) & 0.0003\\
\hspace{3mm}Data Analyst & 12429 (2.42\%) & 12428 (2.42\%) & 0.0000\\
\hspace{3mm}IT Specialist & 21288 (4.15\%) & 21276 (4.15\%) & 0.0001\\
\hspace{3mm}Distribution Specialist & 5443 (1.06\%) & 5445 (1.06\%) & 0.0000\\
\hspace{3mm}Designer & 13879 (2.71\%) & 13890 (2.71\%) & 0.0001\\
Previous seniority, no. (\%) &  &  & \\
\hspace{3mm}1 Entry & 144339 (28.15\%) & 144427 (28.17\%) & 0.0004\\
\hspace{3mm}2 Junior & 157179 (30.66\%) & 157151 (30.66\%) & 0.0001\\
\hspace{3mm}3 Associate & 65857 (12.85\%) & 65829 (12.84\%) & 0.0001\\
\hspace{3mm}4 Manager & 65624 (12.80\%) & 65577 (12.79\%) & 0.0003\\
\hspace{3mm}5 Director & 61335 (11.96\%) & 61327 (11.96\%) & 0.0000\\
\hspace{3mm}6 Executive & 15720 (3.07\%) & 15710 (3.06\%) & 0.0001\\
\hspace{3mm}7 SrExecutive & 2610 (0.51\%) & 2610 (0.51\%) & 0.0000\\
Previous industry (NAICS-2), no. (\%) &  &  & \\
\hspace{3mm}11 & 741 (0.14\%) & 740 (0.14\%) & 0.0000\\
\hspace{3mm}21 & 2661 (0.52\%) & 2660 (0.52\%) & 0.0000\\
\hspace{3mm}22 & 4514 (0.88\%) & 4508 (0.88\%) & 0.0001\\
\hspace{3mm}23 & 6642 (1.30\%) & 6647 (1.30\%) & 0.0001\\
\hspace{3mm}31 & 7328 (1.43\%) & 7322 (1.43\%) & 0.0001\\
\hspace{3mm}32 & 15210 (2.97\%) & 15170 (2.96\%) & 0.0004\\
\hspace{3mm}33 & 39840 (7.77\%) & 39827 (7.77\%) & 0.0001\\
\hspace{3mm}42 & 6953 (1.36\%) & 6948 (1.36\%) & 0.0001\\
\hspace{3mm}44 & 7578 (1.48\%) & 7576 (1.48\%) & 0.0000\\
\hspace{3mm}45 & 14603 (2.85\%) & 14605 (2.85\%) & 0.0000\\
\hspace{3mm}48 & 9821 (1.92\%) & 9822 (1.92\%) & 0.0000\\
\hspace{3mm}49 & 1702 (0.33\%) & 1700 (0.33\%) & 0.0001\\
\hspace{3mm}51 & 95705 (18.67\%) & 95558 (18.64\%) & 0.0007\\
\hspace{3mm}52 & 63837 (12.45\%) & 63887 (12.46\%) & 0.0003\\
\hspace{3mm}53 & 6599 (1.29\%) & 6605 (1.29\%) & 0.0001\\
\hspace{3mm}54 & 86285 (16.83\%) & 86157 (16.81\%) & 0.0006\\
\hspace{3mm}55 & 339 (0.07\%) & 339 (0.07\%) & 0.0000\\
\hspace{3mm}56 & 15659 (3.05\%) & 15662 (3.06\%) & 0.0000\\
\hspace{3mm}61 & 48196 (9.40\%) & 48256 (9.41\%) & 0.0004\\
\hspace{3mm}62 & 22778 (4.44\%) & 22757 (4.44\%) & 0.0002\\
\hspace{3mm}71 & 5937 (1.16\%) & 5944 (1.16\%) & 0.0001\\
\hspace{3mm}72 & 6966 (1.36\%) & 6970 (1.36\%) & 0.0001\\
\hspace{3mm}81 & 15113 (2.95\%) & 15114 (2.95\%) & 0.0000\\
\hspace{3mm}92 & 27559 (5.38\%) & 27758 (5.41\%) & 0.0017\\
\hspace{3mm}99 & 98 (0.02\%) & 98 (0.02\%) & 0.0000\\
Entry year, no. (\%) &  &  & \\
\hspace{3mm}2020 & 41226 (8.04\%) & 41238 (8.04\%) & 0.0001\\
\hspace{3mm}2021 & 117771 (22.97\%) & 117773 (22.97\%) & 0.0000\\
\hspace{3mm}2022 & 113299 (22.10\%) & 113369 (22.12\%) & 0.0004\\
\hspace{3mm}2023 & 125921 (24.56\%) & 125840 (24.55\%) & 0.0003\\
\hspace{3mm}2024 & 114447 (22.32\%) & 114411 (22.32\%) & 0.0001\\
\bottomrule
\end{longtable}
}
\FloatBarrier

\section{Robustness checks supporting the main results}

Table~\ref{tab:si-compare-robustness} compares estimated treatment effects across the baseline specification and three alternative robustness specifications. Across all outcomes, estimated effects remain stable in sign and statistically significant, indicating that the main results are not sensitive to the choice of job-entry window or the operational definition of remote eligibility. As expected, effect sizes are attenuated in the within-individual matched-pairs specification, which absorbs time-invariant individual heterogeneity, but the qualitative patterns and relative ranking across outcomes remain unchanged.

\begin{table}[!htbp]
\centering
\caption{Comparison of overall treatment effects across baseline and robustness specifications.}
\label{tab:si-compare-robustness}
\resizebox{\textwidth}{!}{%
\begin{tabular}[t]{lcccc}
\toprule
\renewcommand{\arraystretch}{1.15}
Outcome & \shortstack{Main\\effect} & \shortstack{Robustness:\\6-month window} & \shortstack{Robustness:\\stricter remote-eligibility\\definition} & \shortstack{Within-individual\\matched pairs}\\
\midrule
Log Wage Growth & 0.040** (0.006) & 0.041** (0.006) & 0.036** (0.006) & 0.021*** (0.001)\\
Upward Seniority Move & 0.020** (0.004) & 0.020** (0.003) & 0.020** (0.003) & 0.009*** (0.001)\\
Employer Prestige Score & -27.179** (5.038) & -26.497** (4.919) & -29.872** (5.403) & -15.991*** (0.252)\\
Log Employer Size & -0.690** (0.122) & -0.693** (0.123) & -1.007** (0.128) & -0.400*** (0.004)\\
Cross-Metro Job Change & 0.019* (0.005) & 0.020* (0.005) & 0.020* (0.005) & 0.004*** (0.001)\\
\bottomrule
\end{tabular}
}

\begin{flushleft}
\footnotesize
\textit{Notes:} The unit of observation is a job-transition record. 
The main specification, as well as the 6-month-window and stricter-definition
specifications, are estimated using Equation~6 and weighted by
inverse probability weights; standard errors are clustered by previous metropolitan area and entry year. 
In the 6-month-window robustness, remote eligibility is measured using employer job postings observed within a six-month window prior to job entry, rather than the baseline three-month window. 
In the stricter-definition robustness, remote eligibility requires a higher
threshold, with the share of remote postings exceeding 0.8 (baseline: 0.5).
The within-individual matched-pairs specification includes matched-pair fixed effects and entry-year fixed effects; standard errors are clustered at the matched-pair level. 
Entries report coefficient estimates with standard errors in parentheses. 
Significance levels are $^{***}p<0.001$, $^{**}p<0.01$, $^{*}p<0.05$, and $^{.}p<0.1$.

\end{flushleft}
\end{table}

\setcounter{section}{4}
\section{Robustness to alternative definitions of upward seniority moves}

Job titles and seniority labels do not have a uniform economic meaning across organizations of different sizes: the same nominal title can correspond to markedly different levels of compensation, responsibility, and organizational scope in large versus small employers. As a result, title-based indicators of upward seniority moves may conflate substantive changes in job rank with differences in title conventions across employers.

This concern is particularly relevant in the context of remote-eligible job transitions, which may expand access to opportunities at smaller or younger employers. To assess whether the upward seniority effects documented in the main analysis reflect substantive changes in seniority rather than nominal title reclassifications, we conduct a robustness analysis based on alternative, employer-size-adjusted definitions of upward seniority moves.

Specifically, we implement a three-step procedure. First, we estimate a calibration model that maps seniority levels and employer size into a common wage--seniority space, net of occupational, geographic, and temporal wage differences. Second, we use this calibrated wage--seniority space to construct employer-size-adjusted measures of upward seniority moves, including both a binary indicator capturing the probability of an upward seniority move and a continuous measure capturing changes in wage-equivalent seniority. Third, we re-estimate the baseline regressions using these adjusted outcomes, holding the sample, weights, fixed effects, and control variables constant.

\subsection{Method: Construction of employer-size-adjusted seniority measures}
This section describes the construction of employer-size-adjusted seniority measures used in the robustness analysis. The objective is to place seniority levels observed at different employers onto a common, economically meaningful scale that is comparable across employers of different sizes.

Our approach anchors seniority to a wage-equivalent scale. Wages provide a natural proxy for the economic content of jobs because they reflect market valuations of skills, responsibility, and organizational scope, and are directly comparable across employers. By calibrating the relationship between seniority and wages as a function of employer size, while controlling for occupational, geographic, and temporal wage differences, we recover a mapping between seniority levels and their wage-equivalent economic meaning that is less sensitive to employer-specific title conventions. Based on this mapping, we construct employer-size-adjusted measures of seniority, including both binary indicators of upward seniority moves and continuous measures of wage-equivalent seniority changes, as described below.

\vspace{0.2cm}
\noindent\textbf{\underline{Step 1. Calibration model.}} 
We estimate a calibration model in which the outcome variable is the logarithm of the wage associated with each job record. Let $i$ index a job record. The explanatory variables include indicators for job seniority, employer size bucket, and their interaction, allowing the wage implications of seniority to vary systematically across employers of different sizes. The model additionally includes fixed effects for occupation, geographic labor market area (metropolitan and non-metropolitan), and calendar year, which absorb systematic wage differences across occupations, locations, and time.

Formally, the calibration model is given by:
\begin{equation}
\log(\text{wage}_{i}) =
f(\text{seniority}_{i}, \text{employer size}_{i}, 
\text{seniority}_{i} \times \text{employer size}_{i})
+ \alpha_{o(i)} + \gamma_{g(i)} + \delta_{t(i)} + \varepsilon_{i},
\end{equation}
where $\alpha_{o(i)}$, $\gamma_{g(i)}$, and $\delta_{t(i)}$ denote fixed effects for occupation, geographic labor market area, and calendar year, respectively. Observations are weighted using inverse-probability weights from the propensity score matching procedure. This calibration model is not used for causal inference; rather, it is employed solely to recover a stable mapping between seniority, employer size, and wage outcomes that defines a common wage--seniority space.

\vspace{0.2cm}
\noindent\textbf{\underline{Step 2. Wage-equivalent seniority mapping.}} 
Using the estimated coefficients from the calibration model, we compute a predicted log wage for each job observation based solely on its seniority level, employer size bucket, and their interaction:
\begin{equation}
\widehat{\log(wage)}^{\mathrm{struct}}_{i} = X_i \hat{\beta},
\end{equation}
where predictions are evaluated excluding all fixed effects. Occupation, geographic labor market area, and calendar year fixed effects are included during estimation to purge confounding wage variation, but are intentionally omitted from prediction in order to recover the structural wage component associated with seniority and employer size alone.

The resulting predicted value represents the wage-equivalent economic content of a given seniority--employer size combination in a common wage--seniority space, abstracting from occupational composition, local labor market conditions, and time-specific shocks.

Aggregating these structural predictions by seniority level and employer size bucket yields a calibrated wage surface (Fig.~\ref{figure_wage_heatmap}), which visualizes how the economic meaning of nominal seniority varies systematically across employers of different sizes.

\vspace{0.2cm}
\noindent\textbf{\underline{Step 3. Definition of upward seniority moves.}} 
In contrast to the title-based indicator defined in Supplementary Methods~S2.2, we define employer-size-adjusted measures of upward seniority moves using the calibrated wage--seniority space. Let $i$ index job observations. For each observation $i$, let $t(i)$ denote the time of the current job spell, and let $t(i)-1$ refer to the worker’s immediately preceding job spell.

The binary employer-size-adjusted upward seniority move indicator records whether the predicted structural wage-equivalent seniority score increases between two consecutive jobs:
\begin{equation}
\label{equ:adjusted_upward_seniority_binary}
\text{UpwardSeniorityMove}_{i}^{\text{bin}} =
\mathbf{1}\!\left\{
\widehat{\log(wage)}^{\mathrm{struct}}_{i,t}
>
\widehat{\log(wage)}^{\mathrm{struct}}_{i,t-1}
\right\}.
\end{equation}

In addition, we define a continuous employer-size-adjusted seniority change measure as the difference in the predicted structural wage-equivalent seniority score between consecutive jobs:
\begin{equation}
\label{equ:adjusted_upward_seniority_cont}
\text{SeniorityChange}_{i}^{\text{cont}} =
\widehat{\log(\text{wage})}_{i,t(i)}^{\text{struct}}
-
\widehat{\log(\text{wage})}_{i,t(i)-1}^{\text{struct}}.
\end{equation}

By construction, the binary measure captures the probability of an upward seniority move in the calibrated wage--seniority space, while the continuous measure captures the magnitude of seniority changes. Both definitions exclude cases in which a higher-sounding title at a smaller employer does not correspond to an increase in the predicted economic content of the job.

\subsection{Descriptive validation: Wage space heatmap} 
\begin{figure}[!htbp] 
	\centering
	\includegraphics[width=\textwidth]{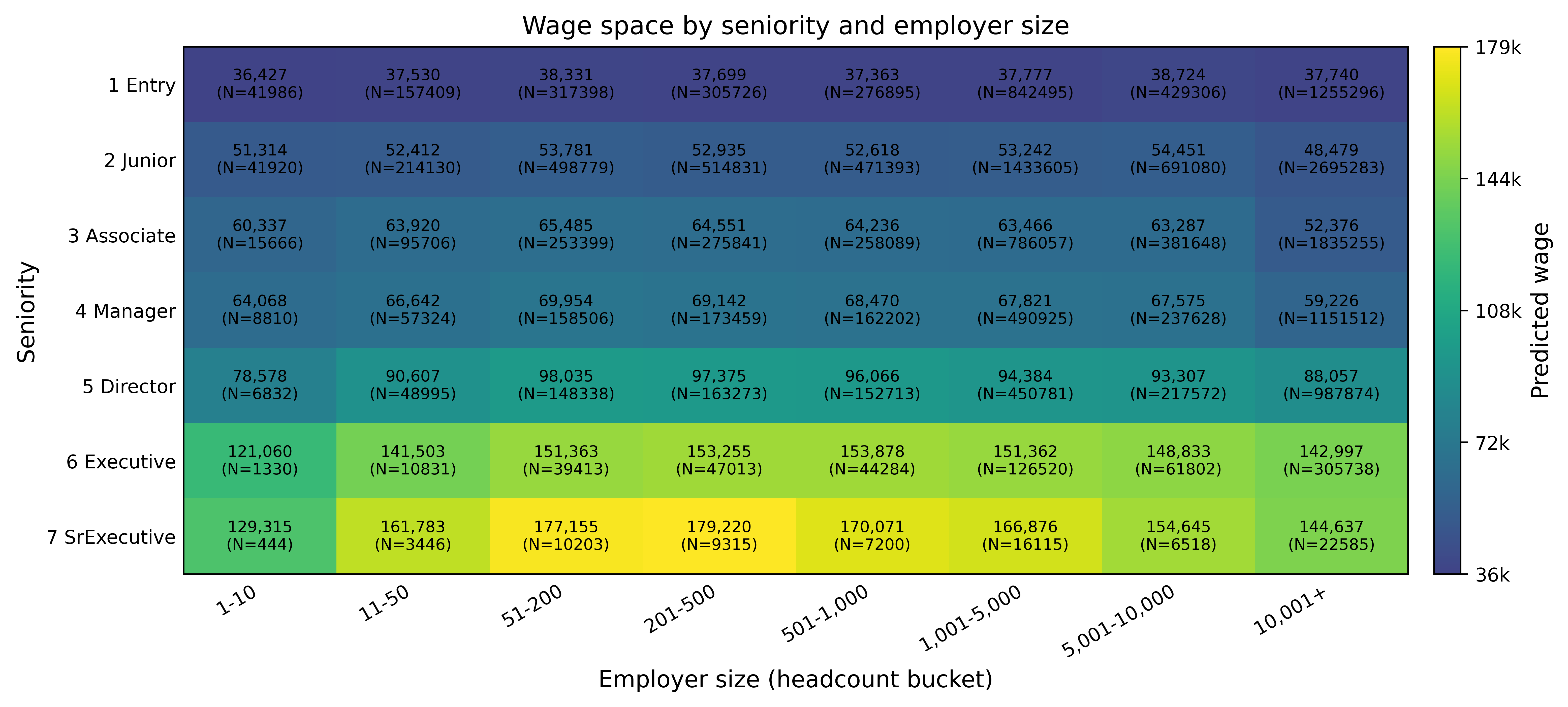} 
	\caption{Wage space by seniority and employer size}
	\label{figure_wage_heatmap} 
\end{figure}

Figure~\ref{figure_wage_heatmap} visualizes the calibrated wage space as a function of job seniority and employer size. Each cell reports the average predicted wage level for a given seniority--employer size combination, obtained from the calibration model described above. Specifically, predicted values are first computed at the individual job level from the structural component of the model—based on seniority, employer size, and their interaction—while excluding occupation, geographic, and year fixed effects, and are then averaged within each seniority--employer size cell. 

Two patterns are immediately apparent. First, wage differences across seniority levels dominate those across employer size. Moving vertically across seniority categories is associated with large, monotonic increases in predicted wages, whereas horizontal movements across employer size buckets within a given seniority level generate comparatively smaller adjustments. This confirms that seniority represents the primary axis of economic content in job titles, while employer size mainly acts as a second-order modifier.

Second, conditional on seniority, predicted wages differ systematically across employer size buckets, although these differences are modest relative to the vertical wage gradients across seniority levels. For example, senior roles at mid-sized and large employers tend to occupy nearby, but not identical, positions in the wage-equivalent space, indicating horizontal shifts in wage levels rather than changes in the seniority hierarchy. We also find that at the upper end of the employer size distribution, larger employers do not always offer higher predicted wages than slightly smaller large employers.

Together, these patterns provide a descriptive validation of two key assumptions underlying the adjusted seniority measure. On the one hand, the dominance of seniority gradients supports the interpretation of reported titles as informative signals of job level. On the other hand, the visible horizontal dispersion within each seniority band highlights why nominal title changes across employers of different sizes may not correspond to equivalent economic advancement. In particular, a move to a higher-sounding title at a smaller employer can leave a worker at a similar—or even lower—position in the wage-equivalent space.

The wage–seniority heatmap therefore motivates the need for employer-size calibration. By anchoring seniority transitions to their predicted wage-equivalent content, the adjusted upward seniority measures used in the analysis filter out purely nominal title inflation and focus on substantive upward mobility in economic terms.

\subsection{Results: Robustness to employer-size-adjusted definitions of upward seniority moves}
We re-estimate the baseline regression specification using employer-size-adjusted measures of upward seniority moves — defined in Equations~S\ref{equ:adjusted_upward_seniority_binary} and S\ref{equ:adjusted_upward_seniority_cont}—as alternative dependent variables. All specifications use the same sample, inverse-probability weights, fixed effects, controls, and clustering as in the baseline analysis. Table~\ref{tab:si-adjusted-promotion-us} reports the estimated coefficients on remote eligibility for the baseline title-based indicator of upward seniority moves, the binary employer-size-adjusted indicator, and the continuous employer-size-adjusted seniority change measure.

Across specifications, the estimated effect of remote eligibility on upward seniority outcomes remains positive and statistically significant (Table~\ref{tab:si-adjusted-promotion-us}). Using the baseline title-based indicator, remote eligibility is associated with a 0.0205 increase in the probability of an upward seniority move. When upward seniority moves are instead defined using the employer-size-adjusted wage--seniority space, the estimated effects remain positive: the coefficient is 0.0328 for the binary employer-size-adjusted indicator and 0.0114 for the continuous employer-size-adjusted seniority change measure. Because the binary indicator captures a probability while the continuous measure reflects changes in wage-equivalent seniority, their magnitudes are not directly comparable. Nevertheless, both adjusted outcomes point to upward movement in the calibrated wage--seniority space among remote-eligible job transitions.

Taken together, these results indicate that the upward seniority advantages associated with remote-eligible job transitions are not solely driven by nominal title differences across employers of different sizes. Remote eligibility is associated with higher probabilities of upward seniority moves even under alternative definitions that require increases in predicted economic content derived from the calibrated wage--seniority space.

\begin{table}[!htbp]
\centering
\caption{\textbf{Robustness: employer-size-adjusted upward seniority measures based on calibrated wage--seniority space.}}
\label{tab:si-adjusted-promotion-us}
\small
\setlength{\tabcolsep}{6pt}
\begin{tabular}{lccc}
\toprule
 & \multicolumn{1}{c}{Estimate} & \multicolumn{1}{c}{Std. error} & \multicolumn{1}{c}{N} \\
\midrule
Upward seniority move (raw indicator) & 0.0205** & (0.0038) & 19,436,473 \\
Adjusted upward seniority move (binary) & 0.0328** & (0.0064) & 19,324,175 \\
Adjusted seniority change (continuous) & 0.0114*** & (0.0013) & 19,324,175 \\
\bottomrule
\end{tabular}
\begin{flushleft}\footnotesize
\textit{Notes:} Each row reports the coefficient on ``Remote-eligible'' from a separate regression estimated on the U.S. overall sample using the same inverse-probability weights as in the main analysis. All specifications include controls for predicted female probability, log previous wage, previous employer prestige, and log previous employer headcount. Fixed effects match the baseline specification, and standard errors are clustered by metropolitan labor market area and entry year. The adjusted outcomes are constructed using a calibration model that maps (seniority $\times$ employer-size bucket) into predicted log wage, defining a common wage--seniority space (see Supplementary Section 5). The binary adjusted outcome captures the probability of an upward seniority move, while the continuous measure captures the magnitude of changes in wage-equivalent seniority. Significance: $^{***}p<0.001$, $^{**}p<0.01$, $^{*}p<0.05$, $^{.}p<0.1$.
\end{flushleft}
\end{table}

\section{Industry heterogeneity in remote eligibility effects}
Remote work adoption differs across industries. To examine whether the effects of remote eligibility depend on industry environments, we conduct two complementary stratified analyses.

\begin{figure}[!htbp] 
	\centering
	\includegraphics[width=\textwidth]{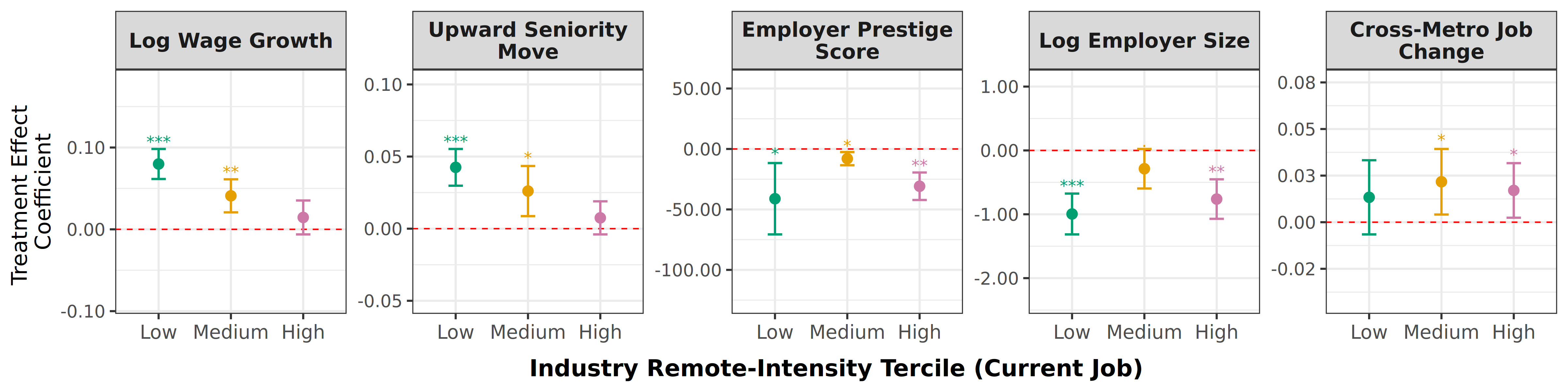} 
	\caption{\textbf{Stratified estimates by industry remote-intensity terciles of the current job.} The figure reports estimated treatment effects of remote eligibility on five career outcomes: log wage growth at job entry, upward seniority moves, employer prestige score, log employer size, and cross–metropolitan job change. Workers are stratified by the remote-work intensity of their current-job industry, measured as the share of remote-eligible occupations within each two-digit NAICS industry. Industries are grouped into terciles (low, medium, and high remote intensity). Points represent regression coefficients from the baseline specification estimated separately within each tercile, and error bars indicate 95\% confidence intervals. Horizontal dashed lines indicate zero effects. Significance levels are $^{***}p<0.001$, $^{**}p<0.01$, $^{*}p<0.05$, and $^{.}p<0.1$.}
	\label{fig:tercile_effects_byNAICS2} 
\end{figure}

First, Supplementary Fig.~\ref{fig:tercile_effects_byNAICS2} stratifies workers by the remote-work intensity of their current-job industry, capturing heterogeneity in the treatment environment. This approach assesses whether the effects of remote eligibility differ systematically across industries with varying baseline levels of remote work. The results indicate that the positive effects on wage growth and upward seniority moves are strongest in low– and medium–remote-intensity industries and attenuated in high–remote-intensity industries, consistent with diminishing marginal returns to remote flexibility in environments where such flexibility is already widespread.

\begin{figure}[!htbp] 
	\centering
	\includegraphics[width=\textwidth]{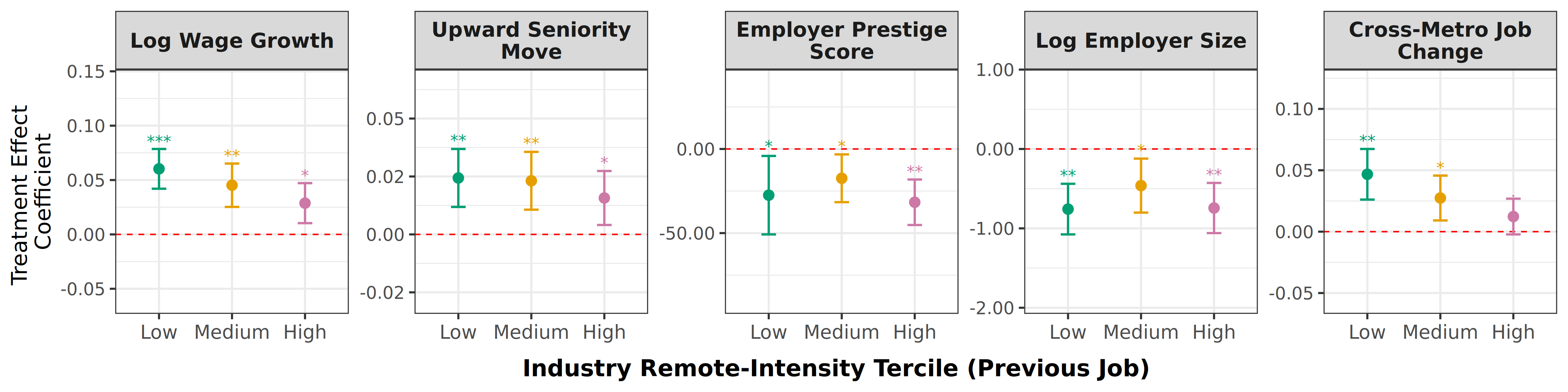} 
	\caption{Stratified estimates by industry remote-intensity terciles of the previous job. The figure reports estimated treatment effects of remote eligibility on five career outcomes: log wage growth at job entry, upward seniority moves, employer prestige score, log employer size, and cross–metropolitan job change. Workers are stratified by the remote-work intensity of their previous-job industry, measured as the share of remote-eligible occupations within each two-digit NAICS industry. Industries are grouped into terciles (low, medium, and high remote intensity). Points represent regression coefficients from the baseline specification estimated separately within each tercile, and error bars indicate 95\% confidence intervals. Horizontal dashed lines indicate zero effects. Significance levels are $^{***}p<0.001$, $^{**}p<0.01$, $^{*}p<0.05$, and $^{.}p<0.1$.}
	\label{fig:tercile_effects_byPrevNAICS2} 
\end{figure}

Second, Supplementary Fig.~\ref{fig:tercile_effects_byPrevNAICS2} implements an origin-based stratification using workers’ previous-job industry. This alternative classification captures heterogeneity across workers coming from different baseline industry contexts, independent of the industry in which the remote-eligible job is located. The estimated effects remain positive across all origin-industry groups, indicating that workers from both high– and low–remote-intensity industries benefit from entering remote-eligible roles.

In sum, these stratifications suggest that the estimated effects of remote eligibility are not confined to a narrow set of high–remote-intensity sectors. Instead, they point to a context-dependent mechanism in which remote eligibility generates the largest marginal gains when it meaningfully expands workers’ effective opportunity sets, while exhibiting weaker effects in industries where remote work is already the norm.

\section{Robustness to occupation-level, time-varying confounding at job transitions}

As discussed in the main text, a remaining limitation is the possibility that time-varying occupational changes at job transitions may partly account for the observed wage effects. In particular, workers may move into higher-paying occupations that are also more remote-feasible, creating a potential source of confounding at the moment of job entry. To further assess the extent to which such occupational upgrading could drive our results, we implement two complementary robustness strategies that operate at different levels of aggregation.

First, Table~\ref{tab:si-within-pairs} exploits within-individual matched pairs and restricts the sample to cases in which workers enter a remote-eligible and a non-remote job within the same occupation. By holding both the individual and occupation constant, this design rules out confounding arising from cross-occupation transitions at job entry. Although coefficient magnitudes are attenuated relative to the baseline specification, the estimated effects remain positive and statistically significant across the main outcomes.

Second, Table~\ref{tab:si-rolek50-yearfe} addresses occupation-level shocks that vary over time in the full regression sample by augmenting the baseline specification with occupation-by-entry-year fixed effects (occupation × entry year). This specification absorbs occupation-level changes that evolve over time, including shifts in labor demand, compensation norms, and the prevalence of remote-feasible work, which may jointly influence remote eligibility and wages. Under this more stringent set of fixed effects, the estimated effect of remote eligibility on log wage growth, as well as on the other non-seniority outcomes, remains positive and statistically significant. 

Taken together, these results suggest that although occupational upgrading may accompany some job transitions, occupation-related changes at job entry are unlikely to explain the estimated effect of remote eligibility.
\setcounter{table}{10}
\begin{table}[!htbp]
\centering
\caption{Within-individual matched-pairs robustness and same-occupation switch.}
\label{tab:si-within-pairs}
\resizebox{\linewidth}{!}{%

\begin{tabular}{lccrrc}
\toprule
Outcome & Within-individual & Same-occupation switch & N (pairs) & N (obs) & Within R$^2$\\
\midrule
Log Wage Growth & 0.02068*** (0.00120) & 0.00504*** (0.00149) & 514,972 / 297,726 & 1,007,531 / 585,581 & 0.0104 / 0.0153\\
Upward Seniority Move & 0.00922*** (0.00107) & 0.00262* (0.00141) & 515,034 / 297,763 & 1,007,808 / 585,759 & 0.0043 / 0.0037\\
Employer Prestige Score & -15.99088*** (0.25151) & -15.20483*** (0.29540) & 515,001 / 297,732 & 1,029,953 / 595,429 & 0.0094 / 0.0102\\
Log Employer Size & -0.40024*** (0.00382) & -0.38932*** (0.00472) & 515,034 / 297,763 & 1,030,068 / 595,526 & 0.0228 / 0.0242\\
Cross-Metro Job Change & 0.00422*** (0.00082) & 0.00251* (0.00106) & 514,066 / 297,301 & 990,104 / 576,174 & 0.0077 / 0.0103\\
\bottomrule
\end{tabular}
}

\begin{flushleft}
\footnotesize
\textit{Notes:} The unit of observation is a job-entry transition record in the matched-pairs sample. Each matched pair contains one remote-eligible and one non-remote job entry experienced by the same individual. All regressions include matched-pair fixed effects and entry-year fixed effects. Standard errors are clustered at the matched-pair level. The \textit{Same-occupation switch} column restricts matched pairs to remote and non-remote entries within the same occupation, matching each remote entry to the temporally closest in-office entry within the same individual and occupation. Columns for $N$ and within $R^2$ report \textit{baseline / same-occupation}. Entries report coefficient estimates with standard errors in parentheses. Significance levels are $^{***}p<0.001$, $^{**}p<0.01$, $^{*}p<0.05$, and $^{.}p<0.1$.
\end{flushleft}
\end{table}

\begin{table}[!htbp]
\centering
\caption{Robustness to occupation-by-entry-year fixed effects. The table compares baseline estimates with a specification absorbing occupation$\times$entry-year fixed effects. Entries report the coefficient on remote eligibility with standard errors in parentheses.}
\label{tab:si-rolek50-yearfe}
\resizebox{\textwidth}{!}{%
\begin{tabular}{lccccc}
\toprule
 & Log wage growth & Upward seniority move & Log employer size & Employer prestige & Cross-metro move \\
\midrule
Baseline & 0.040** & 0.021** & -0.692** & -27.222** & 0.019* \\
 & (0.006) & (0.004) & (0.122) & (5.049) & (0.005) \\
\addlinespace[2pt]
Role$\times$Year FE & 0.023* & 0.009. & -0.649** & -25.415** & 0.020* \\
 & (0.006) & (0.004) & (0.121) & (4.851) & (0.005) \\
\midrule
$N$ (Baseline) & 19433894 & 19436473 & 19436473 & 19434053 & 19436473 \\
$N$ (Role$\times$Year FE) & 19433894 & 19436473 & 19436473 & 19434053 & 19436473 \\
\bottomrule
\end{tabular}
} 
\begin{flushleft}\footnotesize
Notes: All specifications use IPW weights from the baseline matching procedure and two-way clustered standard errors by previous metro area and entry year. Baseline fixed effects include previous seniority, ethnicity, previous occupation, NAICS2, previous metro area, and entry year. The robustness specification replaces occupation and entry-year fixed effects with occupation$\times$entry-year fixed effects while retaining the remaining fixed effects.
\end{flushleft}
\end{table}

\newpage






\end{document}